\documentclass[aps,pre,reprint,onecolumn,groupedaddress,notitlepage]{revtex4-2}

\usepackage{amsmath,amssymb,mathtools}
\usepackage{graphics,subfigure,tabularx}
\usepackage{color}
\usepackage[normalem]{ulem} 
\usepackage{stmaryrd}
\usepackage{txfonts}

\usepackage[colorlinks,hyperindex,linkcolor=blue,citecolor=blue,urlcolor=blue]{hyperref}
\newcommand{\figdraft}{false}                       

\newcommand{\narrow}[1]{\thinmuskip=#1 \medmuskip=1.333\thinmuskip \thickmuskip=1.667\thinmuskip}

\newcommand{\barf}{\bar{f}}
\newcommand{\barg}{\bar{g}}
\newcommand{\barbfm}{\bar{\mathbf{m}}}
\newcommand{\barbfn}{\bar{\mathbf{n}}}

\newcommand{\bfe}{\mathbf{e}}
\newcommand{\bfF}{\mathbf{F}}
\newcommand{\bfg}{\mathbf{g}}
\newcommand{\bfI}{\mathbf{I}}
\newcommand{\bfm}{\mathbf{m}}
\newcommand{\bfM}{\mathbf{M}}
\newcommand{\bfn}{\mathbf{n}}
\newcommand{\bfOmega}{\mathbf{\Omega}}
\newcommand{\bfPi}{\mathbf{\Pi}}
\newcommand{\bfq}{\mathbf{q}}
\newcommand{\bfR}{\mathbf{R}}
\newcommand{\bfS}{\mathbf{S}}
\newcommand{\bfT}{\mathbf{T}}
\newcommand{\bfx}{\mathbf{x}}
\newcommand{\bfX}{\mathbf{X}}
\newcommand{\bfu}{\mathbf{u}}
\newcommand{\bfU}{\mathbf{U}}

\newcommand{\hatF}{\hat{F}}
\newcommand{\hatbfF}{\hat{\mathbf{F}}}

\newcommand{\hatu}{\hat{u}}
\newcommand{\hatbfu}{\hat{\mathbf{u}}}

\newcommand{\itOmega}{\mathit{\Omega}}

\newcommand{\diag}{\mathrm{diag}}
\newcommand{\equ}{\mathrm{eq}}
\newcommand{\ins}{\mathrm{in}}
\newcommand{\MES}{\mathrm{MES}}
\newcommand{\outs}{\mathrm{out}}
\newcommand{\refe}{\mathrm{ref}}

\newcommand{\cm}{\mathrm{cm}}
\newcommand{\g}{\mathrm{g}}
\newcommand{\s}{\mathrm{s}}
\newcommand{\Tr}{\mathrm{T}}

\newcommand{\rhof}{\rho_f^{}}
\newcommand{\rhos}{\rho_s^{}}
\newcommand{\rhor}{\rho_r^{}}

\newcommand{\Gras}{\mathrm{Gr}}
\newcommand{\Reyn}{\mathrm{Re}}
\newcommand{\Prat}{\mathrm{Pr}}

\begin{document}
\title{Volumetric lattice Boltzmann method for thermal particulate flows with conjugate heat transfer}
       
\author{Xiaojie Zhang}
\affiliation{School of Energy Science and Engineering, Central South University, 410083 Changsha, China}
    
\author{Donglei Wang}
\affiliation{School of Energy Science and Engineering, Central South University, 410083 Changsha, China}
    
\author{Qing Li}
\affiliation{School of Energy Science and Engineering, Central South University, 410083 Changsha, China}
    
\author{Rongzong Huang}
\email[Corresponding author: ]{rongzong.huang@csu.edu.cn}
\affiliation{School of Energy Science and Engineering, Central South University, 410083 Changsha, China}

\date{\today}
    
\begin{abstract}
    A volumetric lattice Boltzmann (LB) method is developed for the particle-resolved direct numerical simulation of thermal particulate flows with conjugate heat transfer. This method is devised as a single-domain approach by applying the volumetric interpretation of the LB equation and introducing a solid fraction field to represent the particle. The volumetric LB scheme is employed to enforce the nonslip velocity condition in the solid domain, and a specialized momentum exchange scheme is proposed to calculate the hydrodynamic force and torque acting on the particle. To uniformly solve the temperature field over the entire domain with high numerical fidelity, an energy conservation equation is first derived by reformulating the convection term into a source term. A corresponding LB equation is then devised to automatically achieve the conjugate heat transfer condition and correctly handle the differences in thermophysical properties. Theoretical analysis of this LB equation is also performed to derive the constraints to preserve the numerical fidelity even near the solid-fluid interface. Numerical tests are first performed to validate the present volumetric LB method in various aspects. Then, the sedimentation of a cold particle with conjugate heat transfer in a long channel is investigated. It is found that the sedimentation process can be divided into the accelerating, decelerating, and equilibrium stages. As a further application to dense particulate flows, the sedimentation of 2048 cold particles with conjugate heat transfer in a square cavity is simulated. The particulate Rayleigh-B\'{e}nard convection is successfully captured in this particle-resolved simulation. 
\end{abstract}

\maketitle

\section{Introduction} \label{sec.introduction}
Accurate and efficient simulation of particulate flows is significant in many industrial fields, such as chemical, metallurgy, energy, and microfluidics \citep{Balachandar2010, Maxey2017, Singh2013, DiCarlo2009}. Understanding the transport behaviors of particulate flows has attracted considerable attention in the past decades. Conventional computational fluid dynamics (CFD) methods, such as the finite volume method (FVM) and the finite element method (FEM), have been successfully applied to simulate particulate flows \citep{Apte2009, Shen2022, Pai2009, Capecelatro2013}. Based on the mathematical descriptions of particulate flows, two common strategies have been developed, i.e., the Euler-Euler and Euler-Lagrange strategies \citep{Pai2009, Capecelatro2013, Tenneti2014}. For the Euler-Euler strategy, both the fluid and particles are described by continuous fields in the Eulerian frame. For the Euler-Lagrange strategy, the fluid is described by continuous fields in the Eulerian frame, while the particles are described by discrete points in the Lagrangian frame. The Euler-Euler and Euler-Lagrange strategies provide statistical information on particulate flows. Due to their low computational cost, they can simulate device-scale systems but usually encounter closure problems \citep{Tenneti2014}. As a ``first-principles'' description of particulate flows, the particle-resolved direct numerical simulation (PR DNS) strategy, which is the focus of this work, fully resolves the flow field around every particle and the coupled interaction between fluid and particle. It can thus provide complete information on particulate flows at the price of a high computational cost. The conventional CFD methods for the PR DNS of particulate flows can be grouped into those based on geometrically adapted mesh \citep{Burton2005} and those that employ a fixed Eulerian mesh, such as the immersed boundary (IB) method \citep{Uhlmann2005}. The regeneration of geometrically adapted mesh as the particle moves is quite time-consuming, which worsens when many particles are involved \citep{Nie2010, Chen2013, Cheylan2021}. The methods that employ a fixed Eulerian mesh circumvent the time-consuming mesh regeneration, ensuring its computational efficiency, but require special treatments to satisfy boundary conditions on the particle surface. Nevertheless, due to the fully resolving nature of the PR DNS, the computational cost is always challenging when many particles are involved. As a mesoscopic technique originating from the lattice gas automata, the lattice Boltzmann (LB) method shows many notable merits, like simple algorithm, inherent parallelism, and easy boundary treatment \citep{Higuera1989, Benzi1992}. Thus, since its early development stage, the LB method has been recognized as an efficient and powerful CFD method for the PR DNS of particulate flows \citep{Aidun2010, Succi2015, Ladd1994.A, Ladd1994.B}. A fixed Eulerian mesh is used in the LB method for fluid flows, and various approaches have been proposed to achieve the PR DNS \citep{Ladd1994.A, Ladd1994.B, Noble1998, Feng2004}.

The pioneering work on the LB method for the PR DNS of particulate flows was done by \citet{Ladd1994.A, Ladd1994.B}, which has evolved into a fundamental framework for further development. This approach is based on the physical picture of the halfway bounce back of the density distribution function in the LB method, implying that the solid particle's surface is represented by a series of points located at the middle position between two adjacent lattice nodes. A modified bounce-back scheme is proposed to enforce the nonslip velocity condition at the moving surface of the solid particle. Meanwhile, based on the mesoscopic particle interpretation of the density distribution function in the LB method (a unique interpretation inherited from the lattice gas automata), a momentum exchange scheme is proposed to calculate the hydrodynamic force and torque acting on the solid particle. The zigzag approximation of the particle's surface cannot well preserve the geometric integrity of curved surfaces. Thus, a finer grid is required to satisfy accuracy in real simulations \citep{Mei1999}. To overcome the drawback of the zigzag approximation, various interpolation-based treatments have been proposed for curved surfaces \citep{Mei1999, Bouzidi2001, Filippova1998, Lallemand2003, Li2003}, which can achieve second-order accuracy and ensure a smoother force transition for the moving particle \citep{Chen2013}. As a versatile method for fluid-structure interaction originally proposed by \citet{Peskin1977}, the IB method was first combined with the LB method to achieve the PR DNS of particulate flows by \citet{Feng2004}. In the IB-LB method for particulate flows, the surface of the solid particle is represented and tracked by a series of Lagrangian points, while the velocity field, both inside and outside the particle's surface, is solved on a fixed Eulerian mesh via the LB method. The nonslip velocity condition on the particle's surface is handled by introducing a surface force along it into the LB equation for velocity field. Then, the hydrodynamic force and torque acting on the solid particle can be immediately obtained by integrating this surface force. Various schemes have been proposed to calculate the surface force to achieve better numerical performance, mainly falling into the direct-forcing \citep{Feng2005, Kang2011.A} and feedback-forcing \citep{Feng2004, Kim2007} categories. Based on the volumetric interpretation of the LB equation, \citet{Noble1998} proposed a so-called partially saturated cells (PSC) approach for fluid-structure interaction. The local collision process of the LB equation, without considering the general force term, is modified by introducing an artificial weighting function and an additional collision operator. This modified collision process can be reformulated into the standard collision process with a particular force term \citep{Chen2020}, and thus the hydrodynamic force and torque acting on the solid particle can be directly calculated by summing up this particular force term \citep{Chen2020, Cook2004}. The approach by \citet{Noble1998} fully inherits the merits of the standard LB method, and it has been widely applied to explore dense particulate flows during the past decades \citep{Feng2010, Cui2014, Han2017, Galindo2018}.

Most approaches in the LB method for the PR DNS of particulate flows are proposed for the isothermal situation, and the extensions to thermal particulate flows, where the temperature field should be simultaneously resolved, remain quite limited and essentially open-ended. \citet{Hu2021} proposed interpolation-based treatments for the velocity and temperature conditions on the moving curved boundary to handle particulate flows with thermal convection. Prediction and correction sub-steps, with an enforced iteration, are employed to enhance the numerical accuracy of the boundary condition treatments. \citet{Kang2011.B} extended the direct-forcing IB-LB method for the isothermal situation to thermal particulate flows. An energy source along the particle's surface is introduced to enforce the temperature condition there. \citet{Huang2014} interpreted the surface energy source in the IB method as the latent heat of solid-liquid phase change. Then, they proposed an IB-LB method to simulate solid-liquid phase change problems, where the solid phase is viewed as a solid particle with complex geometry. The unconstrained melting in a circular cylinder is successfully simulated, and the result shows that the free motion of the solid phase can accelerate the melting process. By considering the particle-particle and particle-wall interactions, \citet{Zhang2015} successfully simulated the sedimentations of many isothermal particles (185 and 504 particles) in the cold fluid via the IB-LB method. Similar to the IB method for the velocity condition, several improvements \citep{Wu2017, Tao2021} have been recently proposed to enhance the numerical performance of the IB method for the temperature condition.

It is worth pointing out that the existing LB method for the PR DNS of thermal particulate flows is mainly limited to isothermal particles with a fixed temperature, suggesting that the heat transfer and temperature field are not resolved inside the solid particle. However, thermal particulate flows are typical conjugate heat transfer problems, where the heat transfer inside the solid particle and fluid is coupled in reality. When the Biot number is finite (such as more significant than 0.1 or so), the heat transfer inside the solid particle plays an essential role in the heat transfer process between the solid particle and fluid. Then, the heat transfer and temperature field inside the solid particle should be fully resolved. Under this circumstance, the conjugate heat transfer condition (i.e., the continuity of both the temperature and heat flux) will be encountered at the solid-fluid interface. The temperature and heat flux at the solid-fluid interface are unknown in advance but are part of the solution. In this work, a volumetric LB method is developed for the PR DNS of thermal particulate flows with conjugate heat transfer. This LB method is devised as a single-domain approach, i.e., the velocity and temperature fields are uniformly solved over the entire domain, implying that the merits of the standard LB method can be entirely inherited. The nonslip velocity condition in the solid domain can be strictly ensured, and the hydrodynamic force and torque acting on the solid particle can be accurately calculated. A specialized energy conservation equation is derived, whose form is carefully constructed to be uniformly solved over the entire domain and to preserve the numerical fidelity in the presence of abrupt changes in thermophysical properties across the solid-fluid interface. A corresponding LB equation is then devised to automatically satisfy the conjugate heat transfer condition and correctly handle the differences in thermophysical properties. This LB equation is further analyzed theoretically to derive the constraints to preserve the numerical fidelity near the solid-fluid interface at the discrete level. The remainder of this work is organized as follows. The volumetric LB method is developed in Sec.\ \ref{sec.lbm}, numerical validations and discussions are carried out in Sec.\ \ref{sec.val}, and a brief conclusion is drawn in Sec.\ \ref{sec.con}.

\section{Volumetric lattice Boltzmann method} \label{sec.lbm}
The incompressible Newtonian fluid is considered, and both the viscous heat dissipation and compression work done by pressure are neglected. The governing equations of the thermal fluid flows can be written as \citep{He1998} 
\begin{subequations}\label{eq.nse}
    \begin{gather}
        \label{eq.nse.rho}
        \dfrac{\partial \rho}{\partial t} + \nabla \cdot (\rho \bfu) = 0,
        \\
        \label{eq.nse.rhou}
        \dfrac{\partial (\rho \bfu)}{\partial t} + \nabla \cdot (\rho \bfu \bfu) = -\nabla p + \bfF + \nabla \cdot \bfPi, 
        \\
        \label{eq.nse.rhoe}
        \dfrac{\partial (\rho \epsilon)}{\partial t} + \nabla \cdot (\rho \epsilon \bfu) = \nabla \cdot (\lambda \nabla T), 
    \end{gather}
\end{subequations}
where $\rho$ is the density, $\bfu$ is the velocity, $p$ is the pressure, $\bfF$ is the body force like gravity, $\bfPi$ is the Newtonian viscous stress, $\epsilon$ is the internal energy, $T$ is the temperature, and $\lambda$ is the heat conductivity. Equation (\ref{eq.nse}) can be solved by the well-established double-distribution-function (DDF) LB method \citep{Huang2016, He1998, Huang2019}.

\begin{figure}[tbp]
    \centering
    \includegraphics[scale=1.0,draft=\figdraft]{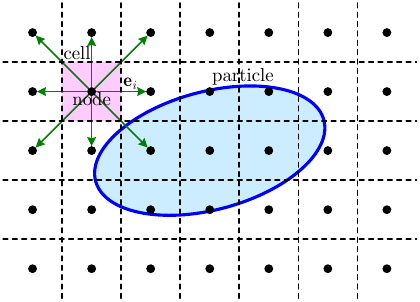}
    \caption[]{Illustration of the volumetric interpretation of the LB equation, where the point denotes the lattice node, the dashed line denotes the lattice grid with the forming dashed box the lattice cell, the arrow denotes the lattice discrete velocity, and the elliptic denotes a solid particle.}
    \label{fig.01}
\end{figure}

For the thermal particulate flows, the solid particle is immersed in the fluid, and the thermophysical properties (such as the density, specific heat, and heat conductivity) of the solid usually differ from those of the fluid. On the particle's surface, the nonslip velocity and conjugate heat transfer conditions are encountered \citep{Chen2000}. In this work, a volumetric LB method is developed for the PR DNS of thermal particulate flows with conjugate heat transfer. The volumetric interpretation of the LB equation is applied \citep{Chen2006}, and thus the lattice node is located at the center of the lattice cell, as illustrated by Fig.\ \ref{fig.01}. With this volumetric interpretation, a solid fraction $f_s$, defined as the volume percentage of the solid part within the lattice cell, is introduced here, and the fluid, solid, and interface lattice cells are denoted by $f_s = 0$, $f_s = 1$, and $0 < f_s < 1$, respectively. Accordingly, the solid particle can be represented by the solid fraction field. Under this scenario, the present volumetric LB method for thermal particulate flows can be devised as a single-domain approach (i.e., all three kinds of lattice cells uniformly participate in the computations without unique treatments), and it should degenerate into the conventional DDF LB method for thermal fluid flows when $f_s = 0$. The two-dimensional nine-velocity (D2Q9) discrete velocity set is considered in this work, and the extension to the three-dimensional version is straightforward. The nine discrete velocities are given as \citep{Qian1992}
\begin{equation}
    \bfe_i = \begin{dcases}
        c \left( 0 ,\, 0 \right) ^\Tr, & i = 0, \\
        c \left( \cos[(i-1)\pi/2] ,\, \sin[(i-1)\pi/2] \right) ^\Tr, & i = 1, 2, 3, 4, \\
        \sqrt{2}c \left( \cos[(2i-1)\pi/4] ,\, \sin[(2i-1)\pi/4] \right) ^\Tr, & i = 5, 6, 7, 8, \\
    \end{dcases}
\end{equation}
where $c = \delta_x / \delta_t$ is the lattice speed, $\delta_x$ and $\delta_t$ are the lattice spacing and the time step, respectively.

\subsection{LB equation for velocity field}\label{sec.lbm.velocity}
As a single-domain approach, the present LB equation for velocity field is devised to be uniformly solved over the entire domain, and the pivotal point is to ensure the nonslip velocity condition in the domain occupied by the solid particle, not only at the solid-fluid interface but also within the solid particle. Solving the velocity field in the fluid domain requires only the nonslip velocity condition at the solid-fluid interface. However, to uniformly solve the temperature field over the entire domain and automatically satisfy the conjugate heat transfer condition at the solid-fluid interface (see Sec.\ \ref{sec.lbm.temperature}), as well as to precisely calculate the hydrodynamic force and torque acting on the solid particle (see Sec.\ \ref{sec.lbm.particle}), the nonslip velocity condition within the solid particle is also needed. For this purpose, the volumetric LB scheme, initially proposed for the solid-liquid phase change \citep{Huang2016}, is extended to particulate flows here. The multiple-relaxation-time (MRT) LB equation for the density distribution function $f_i (\bfx, t)$ can be expressed as \citep{Huang2016} 
\begin{subequations}\label{eq.lbe.rho}
    \begin{gather}
        \label{eq.lbe.rho.col}
        \barbfm (\bfx, t) = \bfm (\bfx, t) - \bfS \left[ \bfm (\bfx, t) - \bfm^\equ (\bfx, t) \right] + \delta_t \left( \bfI - \dfrac{\bfS}{2} \right) \bfF_m (\bfx, t) ,
        \\
        \label{eq.lbe.rho.str}
        f_i^\ast (\bfx + \bfe_i \delta_t, t + \delta_t) = \barf_i (\bfx, t) , 
    \end{gather}
\end{subequations}
where $\bfm = \bfM \left( f_0, f_1, \cdots, f_8 \right) ^\Tr$ is the moment of the density distribution function with $\bfM$ the dimensionless orthogonal transformation matrix, $\bfI$ is the $9 \times 9$ unit matrix, the overbar ``--'' denotes the post-collision state, and the superscript ``$\ast$'' means that the nonslip velocity condition in the solid domain is not yet considered, which will be further discussed below. Equations (\ref{eq.lbe.rho.col}) and (\ref{eq.lbe.rho.str}) are known as the local collision and linear streaming processes in the LB method, respectively, and the present collision process for particulate flows remains identical to the conventional one for fluid flows. For the D2Q9 discrete velocity set, the dimensionless orthogonal transformation matrix $\bfM$ can be chosen as \citep{Huang2015, Lallemand2000} 
\begin{equation}
    \bfM = \begin{bmatrix}
        1& 1& 1& 1& 1& 1& 1& 1& 1 \\
        -4&-1&-1&-1&-1& 2& 2& 2& 2 \\
        4&-2&-2&-2&-2& 1& 1& 1& 1 \\
        0& 1& 0&-1& 0& 1&-1&-1& 1 \\
        0&-2& 0& 2& 0& 1&-1&-1& 1 \\
        0& 0& 1& 0&-1& 1& 1&-1&-1 \\
        0& 0&-2& 0& 2& 1& 1&-1&-1 \\
        0& 1&-1& 1&-1& 0& 0& 0& 0 \\
        0& 0& 0& 0& 0& 1&-1& 1&-1
    \end{bmatrix}.
\end{equation}
The diagonal relaxation matrix $\bfS$ in the moment space is given as \citep{Lallemand2000}
\begin{equation}
    \bfS = \diag (s_0, s_e, s_\varepsilon, s_j, s_q, s_j, s_q, s_p, s_p) . 
\end{equation} 
The equilibrium moment, defined as $\bfm^\equ = \bfM \left( f_0^\equ , f_1^\equ , \cdots, f_8^\equ \right) ^\Tr$ with $f_i^\equ$ the equilibrium density distribution function, is given as \citep{Lallemand2000}
\begin{equation}
    \bfm^\equ = \left[ \rho ,\; -2\rho + 3\rho |\hatbfu|^2 ,\; \rho - 3\rho |\hatbfu|^2 ,\; \rho \hatu_x ,\; -\rho \hatu_x ,\; \rho \hatu_y ,\; -\rho \hatu_y ,\; \rho \big( \hatu_x^2 - \hatu_y^2 \big) ,\; \rho \hatu_x \hatu_y \right] ^\Tr, 
\end{equation}
where $\hatbfu = (\hatu_x ,\, \hatu_y)^\Tr = \bfu / c$ with $\bfu$ the macroscopic velocity. The discrete force term $\bfF_m$ in the moment space is given as \citep{Guo2002, Guo2008} 
\begin{equation}
    \bfF_m = \left[ 0 ,\; 6 \hatbfF \cdot \hatbfu ,\; -6 \hatbfF \cdot \hatbfu ,\; \hatF_x ,\; -\hatF_x ,\; \hatF_y ,\; -\hatF_y ,\; 2 \big( \hatF_x \hatu_x - \hatF_y \hatu_y \big) ,\; \hatF_x \hatu_y + \hatF_y \hatu_x \right] ^\Tr, 
\end{equation}
where $\hatbfF = (\hatF_x ,\, \hatF_y)^\Tr = \bfF / c$ with $\bfF$ the body force (such as gravity or buoyancy). Note that there is no need to restrict $c = 1$ for the present MRT LB equation as the rescaled moment is adopted \citep{Huang2015}.

The density distribution function $f_i^\ast (\bfx, t + \delta_t)$ evolved according to the LB equation [i.e., Eq.\ (\ref{eq.lbe.rho})] is temporary without considering the nonslip velocity condition in the solid domain. Based on the volumetric interpretation (see Fig.\ \ref{fig.01}), each lattice cell consists of fluid and solid fractions. For the fluid fraction $1 - f_s (\bfx, t + \delta_t)$, the density distribution function is the temporal one given by Eq.\ (\ref{eq.lbe.rho}). In contrast, for the solid fraction $f_s (\bfx, t + \delta_t)$, the density distribution function is assumed to be at the equilibrium state \citep{Huang2016}. Therefore, the desired density distribution function $f_i (\bfx, t + \delta_t)$ can be locally obtained by linear interpolation as follows:
\begin{equation}\label{eq.fi}
    f_i = (1-f_s) f_i^\ast + f_s f_i^\equ (\rho, \bfu_s) ,
\end{equation}
where $\rho (\bfx, t + \delta_t)$ is the density given by $\rho = \sum\nolimits_{i=0}^{8} f_i^\ast$, $\bfu_s (\bfx, t + \delta_t)$ is the velocity of the solid fraction determined by the rigid body motion of the solid particle [see Eq.\ (\ref{eq.us})], and the solid fraction $f_s (\bfx, t + \delta_t)$ is directly determined by the position of the solid particle, which is updated via particle dynamics in Sec.\ \ref{sec.lbm.particle}. With the desired density distribution function $f_i (\bfx, t + \delta_t)$, the macroscopic density $\rho (\bfx, t + \delta_t)$ and velocity $\bfu (\bfx, t + \delta_t)$ can be uniformly calculated over the entire domain as
\begin{equation}\label{eq.rho.rhou}
    \rho = \sum\limits_{i=0}^{8} f_i, \qquad  \rho \bfu = \sum\limits_{i=0}^{8} \bfe_i f_i + \dfrac{\delta_t}{2} \bfF .
\end{equation}
Based on Eqs.\ (\ref{eq.fi}) and (\ref{eq.rho.rhou}), there is $\rho = \sum\nolimits_{i=0}^{8} f_i = \sum\nolimits_{i=0}^{8} f_i^\ast$, implying that the local mass conservation can be strictly satisfied. Meanwhile, for the solid cell with $f_s = 1$, there is $\bfu = \bfu_s$, implying that the nonslip velocity condition in the solid domain is strictly ensured. For the fluid cell with $f_s = 0$, Eqs.\ (\ref{eq.fi}) and (\ref{eq.rho.rhou}), together with Eq.\ (\ref{eq.lbe.rho}), precisely degenerate into the standard LB method for fluid flows. Consequently, the kinematic and bulk viscosities of the fluid are recovered as $\nu = c_s^2 \delta_t ( s_p^{-1} - 0.5 )$ and $\varsigma = c_s^2 \delta_t ( s_e^{-1} - 0.5 )$, respectively, and the pressure is determined by $p = \rho c_s^2$ \citep{Lallemand2000}. Here, $c_s = c / \sqrt{3}$ is the lattice sound speed, and the dimensionless relaxation time for the density distribution function can be defined as $\tau_f = 1 / s_p$. Furthermore, as can be seen from Eqs.\ (\ref{eq.lbe.rho}), (\ref{eq.fi}), and (\ref{eq.rho.rhou}), the nonslip velocity condition in the solid domain is implemented by modifying the linear streaming process while keeping the local collision process unchanged. Therefore, the present LB equation for velocity field for particulate flows is expected to apply when the fluid itself is subjected to body force, considering the force term is implemented by the local collision process [see Eq.\ (\ref{eq.lbe.rho.col})]. Numerical tests will be carried out to demonstrate this point in Sec.\ \ref{sec.sed.iso}. Consequently, the present LB equation for velocity field is capable of handling thermal particulate flows where buoyancy, at least, should be considered. Before proceeding further, it is worth pointing out that the solid and interface lattice cells with $f_s > 0$ participate in all the computations (including the collision and streaming processes) just as the fluid lattice cell with $f_s = 0$. Therefore, the solid fraction of these lattice cells can be interpreted as being filled with \textit{numerical fluid}, which does not take effect in solving the velocity field but should be seriously considered in calculating the hydrodynamic force and torque acting on the solid particle (see Sec.\ \ref{sec.lbm.particle}).

\subsection{LB equation for temperature field}\label{sec.lbm.temperature}
For the PR DNS of thermal particulate flows with conjugate heat transfer, both the temperature fields in the fluid and solid domains should be fully resolved, and the conjugate heat transfer condition at the solid-fluid interface (i.e., the continuity of both the temperature and heat flux) should be strictly satisfied. Since the nonslip velocity condition in the solid domain can be enforced by the above LB equation for velocity field, the energy conservation equation [i.e., Eq.\ (\ref{eq.nse.rhoe})] can then be directly solved over the entire domain, and the conjugate heat transfer condition at the solid-fluid interface can be automatically satisfied without any additional treatment. Under this scenario, the internal energy in Eq.\ (\ref{eq.nse.rhoe}) is uniformly expressed as $\epsilon = c_v T$, where the specific heat $c_v$ denotes $c_{v,f}$ in the fluid domain and $c_{v,s}$ in the solid domain. This work focuses on incompressible flows, implying that the fluid density $\rho$ keeps constant and the velocity field is divergence-free (i.e., $\nabla \cdot \bfu = 0$) from the fluid mechanics perspective. It is worth noting that the mass conservation equation recovered by the LB equation in Sec.\ \ref{sec.lbm.velocity} via the Chapman-Enskog analysis is $\partial \rho \big/ \partial t + \nabla \cdot (\rho \bfu) = 0$. However, this LB equation also requires the common low Mach number assumption in the LB method, implying it is only suitable for low-speed flows with neglectable density variation. Since the solid is also incompressible, the density $\rho$ and specific heat $c_v$ remain constant in both the fluid and solid domains, respectively. Therefore, Eq.\ (\ref{eq.nse.rhoe}) can be rewritten as
\begin{equation}\label{eq.nse.e}
    \dfrac{\partial \epsilon}{\partial t} + \nabla \cdot (\epsilon \bfu) = \nabla \cdot \left( \dfrac{\lambda}{\rhof} \nabla T \right) ,
\end{equation}
where $\rhof$ denotes the exact density of the fluid, and the internal energy can still be uniformly expressed as $\epsilon = c_v T$, with $c_v$ being $c_{v,f}$ in the fluid domain but $\rhos c_{v,s} / \rhof$ in the solid domain. Here, $\rhos$ denotes the exact density of the solid, and $\rhos c_{v,s} / \rhof$ can be viewed as a \textit{pseudo specific heat of the solid}. To simplify the notation, the exact specific heat of the solid will not be referred to, and the \textit{pseudo specific heat of the solid} $\rhos c_{v,s} / \rhof$ is denoted by the solid specific heat $c_{v,s}$ in the following. Such a mathematical treatment does not introduce any limitations in real applications. With the present volumetric interpretation (see Fig.\ \ref{fig.01}), the specific heat $c_v$ and heat conductivity $\lambda$ in Eq.\ (\ref{eq.nse.e}) can be uniformly expressed as 
\begin{subequations}
    \begin{gather}
        c_v = (1-f_s) c_{v,f} + f_s c_{v,s} ,
        \\
        \lambda = (1-f_s) \lambda_f + f_s \lambda_s ,
    \end{gather}
\end{subequations}
where $\lambda_f$ and $\lambda_s$ denote the heat conductivities of the fluid and solid, respectively. In real applications, the thermophysical properties of the solid could differ from those of the fluid, implying that both $c_v$ and $\lambda$ in Eq.\ (\ref{eq.nse.e}) could abruptly change across the solid-fluid interface. These abrupt changes pose a significant challenge to the numerical fidelity of modeling and simulation in the framework of the single-domain approach.

Equation (\ref{eq.nse.e}) can be directly solved by the total enthalpy-based LB method for solid-liquid phase change by neglecting the latent enthalpy \citep{Huang2015, Huang2016}. For conjugate heat transfer problems where the solid is stationary, the specific heat $c_v$ and heat conductivity $\lambda$ in Eq.\ (\ref{eq.nse.e}) only vary with space but not time and satisfying numerical results can be obtained (see \citet{Huang2015} for details). However, for the thermal particulate flows focused in this work, the solid particle moves with the fluid, leading to $c_v$ and $\lambda$ in Eq.\ (\ref{eq.nse.e}) steeply varying with both space and time, and a significant numerical error will be induced by the steep variation of $c_v$ near the solid-fluid interface based on our numerical tests. To circumvent such numerical difficulty, we reformulate the convection term in Eq.\ (\ref{eq.nse.e}) as follows:
\begin{equation}\label{eq.convt}
    \begin{split}
        \nabla \cdot (\epsilon \bfu) \equiv \nabla \cdot (c_v T \bfu) &= c_v \bfu \cdot \nabla T + T \nabla \cdot (c_v \bfu) \\
        &=c_v \bfu \cdot \nabla T + T \nabla \cdot \Big\{ [(1-f_s) c_{v,f} + f_s c_{v,s}] \bfu \Big\} \\
        &=c_v \bfu \cdot \nabla T + (c_{v,s} - c_{v,f}) T \nabla \cdot (f_s \bfu) + c_{v,f} T \nabla \cdot \bfu .
    \end{split}
\end{equation}
In the fluid domain, the divergence of the velocity vanishes from the fluid mechanics perspective as this work focuses on incompressible flows. In the solid domain, the velocity field satisfies the rigid body motion of the solid particle, implying that the divergence of the velocity also vanishes. Therefore, there is $\nabla \cdot \bfu = 0$ over the entire domain, and thus the last term $c_{v,f} T \nabla \cdot \bfu$ in Eq.\ (\ref{eq.convt}) can be uniformly neglected. To simplify the second term $(c_{v,s} - c_{v,f}) T \nabla \cdot (f_s \bfu)$ in Eq.\ (\ref{eq.convt}), a governing equation for the solid fraction $f_s$ should first be established. For this purpose, an arbitrary material element $\itOmega (t)$ is considered here, and the total solid fraction within $\itOmega (t)$ can be expressed as $\int\nolimits_{\itOmega (t)} f_s d V$. In reality, the solid-fluid interface always remains sharp rather than diffuses as the solid particle moves, implying that the total solid fraction $\int\nolimits_{\itOmega (t)} f_s d V$ should remain constant, i.e., there is
\begin{equation}\label{eq.fs.con}
    \dfrac{D}{D t} \int\nolimits_{\itOmega (t)} f_s d V = 0 ,
\end{equation} 
where $D / D t = \partial / \partial t + \bfu \cdot \nabla$ represents the material derivative. From the fluid mechanics perspective, the velocity divergence can be interpreted as the relative volume expansion rate of an infinitesimal element, i.e., there is $\nabla \cdot \bfu = [ D (d V) / D t ] \big/ d V$. Using this relation, the left-hand side of Eq.\ (\ref{eq.fs.con}) can be formulated as
\begin{equation}
    \begin{split}
        \dfrac{D}{D t} \int_{\itOmega (t)} f_s d V  &= \int_{\itOmega (t)} \dfrac{D f_s}{D t} d V + \int_{\itOmega (t)} f_s \dfrac{ D (d V) }{ D t } \\
        &= \int_{\itOmega (t)} \left( \dfrac{\partial f_s}{\partial t} + \bfu \cdot \nabla f_s \right) d V  + \int_{\itOmega (t)} f_s (\nabla \cdot \bfu) d V \\
        &= \int_{\itOmega (t)} \left[ \dfrac{\partial f_s}{\partial t} + \nabla \cdot (f_s \bfu) \right] d V .
    \end{split}
\end{equation}
Substituting this expression into Eq.\ (\ref{eq.fs.con}) and considering the arbitrariness of the material element $\itOmega (t)$, Eq.\ (\ref{eq.fs.con}) immediately leads to
\begin{equation}\label{eq.fs}
    \dfrac{\partial f_s}{\partial t} + \nabla \cdot (f_s \bfu) = 0, 
\end{equation}
which suggests that the solid fraction $f_s$ can be viewed as a scalar variable transported with the fluid and governed by a pure convection equation. With this governing equation for $f_s$, the second term $(c_{v,s} - c_{v,f}) T \nabla \cdot (f_s \bfu)$ in Eq.\ (\ref{eq.convt}) can be simplified as $- (c_{v,s} - c_{v,f}) T \partial f_s \big/ \partial t$. Based on the above discussions, the energy conservation equation [i.e., Eq.\ (\ref{eq.nse.e})] can finally be reformulated as
\begin{equation}\label{eq.nse.e.final}
    \dfrac{\partial \epsilon}{\partial t} = \nabla \cdot \left( \dfrac{\lambda}{\rhof} \nabla T \right) - c_v \bfu \cdot \nabla T + (c_{v,s} - c_{v,f}) T \dfrac{\partial f_s}{\partial t} ,
\end{equation}
which can be viewed as a pure diffusion equation with a source term $q_c^{} = - c_v \bfu \cdot \nabla T + (c_{v,s} - c_{v,f}) T \partial f_s \big/ \partial t$. Here, it is worth emphasizing that the governing equation for $f_s$ [i.e., Eq.\ (\ref{eq.fs})] is only used to simplify the second term in Eq.\ (\ref{eq.convt}). In real simulations, the solid fraction $f_s$ is directly determined by the position of the solid particle, and then the time derivative $\partial f_s \big/ \partial t$ can be calculated, which will be further discussed. Here, it is worth pointing out that the low Mach number assumption in the LB method implies that the moving velocity of the solid particle is rather small compared with the lattice speed $c = \delta_x \big/ \delta_t$, suggesting that the solid particle won't jump a couple of lattices during one time step. Such a scenario ensures the calculating accuracy of $\partial f_s \big/ \partial t$ at each lattice cell.

To solve Eq.\ (\ref{eq.nse.e.final}) via the LB method, a distribution function for $\epsilon$, i.e., the internal energy distribution function $g_i^{} (\bfx, t)$, is introduced here. The MRT LB equation for $g_i^{} (\bfx, t)$ can be expressed as 
\begin{subequations}\label{eq.lbe.e}
    \begin{gather}
        \label{eq.lbe.e.col}
        \barbfn (\bfx, t) = \bfn (\bfx, t) - \bfR \left[ \bfn (\bfx, t) - \bfn^\equ (\bfx, t) \right] + \delta_t \left( \bfI - \dfrac{\bfR}{2} \right) \bfq_m^{} (\bfx, t) ,
        \\
        \label{eq.lbe.e.str}
        g_i^{} (\bfx + \bfe_i \delta_t, t + \delta_t) = \barg_i^{} (\bfx, t) , 
    \end{gather}
\end{subequations}
where $\bfn = \bfM ( g_0^{}, g_1^{}, \cdots , g_8^{} ) ^\Tr$ is the moment of the internal energy distribution function, and $\bfR$ is the diagonal relaxation matrix in the moment space which is given as
\begin{equation}
    \bfR = \diag (\sigma_0, \sigma_e, \sigma_\varepsilon, \sigma_j, \sigma_q, \sigma_j, \sigma_q, \sigma_e, \sigma_e) . 
\end{equation} 
Following the work by \citet{Huang2015}, the equilibrium moment, defined as $\bfn^\equ = \bfM \left( g_0^\equ , g_1^\equ , \cdots, g_8^\equ \right) ^\Tr$ with $g_i^\equ$ the equilibrium distribution function for the internal energy, is devised as
\begin{equation}
    \bfn^\equ = \left[ \epsilon ,\; -4\epsilon + (4+\alpha_1) c_{v,\refe} T ,\; 4\epsilon - (4-\alpha_2) c_{v,\refe} T ,\; 0 ,\; 0 ,\; 0 ,\; 0 ,\; 0 ,\; 0 \right] ^\Tr ,
\end{equation}
where $\alpha_1$ and $\alpha_2$ are free parameters, and $c_{v,\refe}$ is a reference specific heat introduced to correctly handle the differences in thermophysical properties between the fluid and solid. The value of $c_{v,\refe}$ can be arbitrarily chosen but should be constant over the entire domain \citep{Huang2015}. Without loss of generality, it is set to the harmonic mean of the fluid and solid specific heats, i.e., $c_{v,\refe} = 2 c_{v,f} c_{v,s} \big/ (c_{v,f} + c_{v,s})$, throughout this work. The discrete source term $\bfq_m^{}$ in the moment space is given as \citep{Huang2015.CDE}
\begin{equation}
    \bfq_m^{} = \left[ q_c^{} ,\; \beta_1 q_c^{} ,\; \beta_2 q_c^{} ,\; 0 ,\; 0 ,\; 0 ,\; 0 ,\; 0 ,\; 0 \right] ^\Tr ,
\end{equation}
where $\beta_1$ should be fixed at $-4$ to eliminate the significant numerical error induced by the steep variation of $c_v$ near the solid-fluid interface (see Appendix \ref{app.analysis}), and $\beta_2$ is a free parameter. The macroscopic internal energy $\epsilon (\bfx, t + \delta_t)$ is defined as
\begin{equation}\label{eq.e}
    \epsilon = \sum\limits_{i=0}^{8} g_i^{} + \dfrac{\delta_t}{2} q_c^{} ,
\end{equation}
and then the temperature $T (\bfx, t+\delta_t)$ can be obtained via the thermodynamic relation $T = \epsilon / c_v$. Through the Chapman-Enskog analysis, the targeted energy conservation equation [i.e., Eq.\ (\ref{eq.nse.e.final})] can be correctly recovered, where the heat conductivity satisfies
\begin{equation}\label{eq.lambda}
    \dfrac{\lambda}{\rhof} = \dfrac{4+\alpha_1}{6} c_{v,\refe} c^2 \delta_t \left( \sigma_j^{-1} - 0.5 \right) .
\end{equation}
Here, the dimensionless relaxation time for the internal energy distribution function can be defined as $\tau_g = 1 / \sigma_j$. In real simulations, the above free parameters $\alpha_1$, $\alpha_2$, and $\beta_2$ can be adjusted to achieve better numerical stability and accuracy, and they are set to $\alpha_1 = -2$, $\alpha_2 = 1$, and $\beta_2 = 4$, respectively, in this work.

Before proceeding further, some discussion on computing the convection term in the form of a source term, i.e., the $q_c^{} = -c_v \bfu \cdot \nabla T + (c_{v,s} - c_{v,f}) T \partial f_s \big/ \partial t$ in Eq.\ (\ref{eq.nse.e.final}), is useful. Since the thermal particulate flows with conjugate heat transfer are usually transient problems, it is necessary to ensure that the source term $q_c^{}$ is also at position $\bfx$ and time $t + \delta_t$ when computing the macroscopic variables $\epsilon (\bfx, t+\delta_t)$ and $T (\bfx, t+\delta_t)$ at the next time step [see Eq.\ (\ref{eq.e})]. For this purpose, the temperature gradient $\nabla T$ at position $\bfx$ and time $t+\delta_t$, denoted by $\big[ \nabla T \big] (\bfx, t+\delta_t)$, is computed by the following local scheme
\begin{equation}
    \big[ \nabla T \big] (\bfx, t + \delta_t) = - \dfrac{6}{4 + \alpha_1} \dfrac{1}{c_{v,\refe} \delta_t} \dfrac{\sum\nolimits_{i=0}^{8} \bfe_i g_i^{} (\bfx, t+\delta_t)}{ \tau_g (\bfx, t+\delta_t) } ,
\end{equation}
which is derived from the Chapman-Enskog analysis and widely accepted in the LB method \citep{Huang2015.CDE}. Here, $g_i^{} (\bfx, t+\delta_t)$ is determined by the LB equation, and $\tau_g (\bfx, t+\delta_t)$ is determined by Eq.\ (\ref{eq.lambda}). The time derivative of the solid fraction $\partial f_s \big/ \partial t$ at position $\bfx$ and time $t+\delta_t$, denoted by $\big[ \partial f_s \big/ \partial t\big] (\bfx, t+\delta_t)$, is computed as follows:
\begin{equation}\label{eq.dfs_dt}
    \bigg[ \dfrac{\partial f_s}{\partial t} \bigg] (\bfx, t + \delta_t) = 2 \dfrac{f_s (\bfx, t + \delta_t) - f_s (\bfx, t) }{\delta_t} - \bigg[ \dfrac{\partial f_s}{\partial t} \bigg] (\bfx,t) ,
\end{equation}
where $f_s (\bfx, t+\delta_t)$ is determined by the rigid body motion of the solid particle, and $\big[ f_s (\bfx, t+\delta_t) - f_s (\bfx, t)\big] \big/ \delta_t$ can be interpreted as a central difference for $\partial f_s \big/ \partial t$ at position $\bfx$ and time $t + \delta_t / 2$. Therefore, Eq.\ (\ref{eq.dfs_dt}) integrates the central difference and linear extrapolation schemes, which helps to preserve the numerical fidelity near the solid-fluid interface (see Appendix \ref{app.analysis}). Here, it is worth emphasizing that $\big[ \partial f_s \big/ \partial t\big] (\bfx, t+\delta_t)$ cannot be simply computed by the backward difference $\big[ f_s (\bfx, t+\delta_t) - f_s (\bfx, t)\big] \big/ \delta_t$. Otherwise, a significant numerical error will inevitably be introduced near the solid-fluid interface. Note that $T (\bfx, t+\delta_t)$ is also involved in $q_c^{} (\bfx, t+\delta_t)$. Therefore, $T (\bfx, t+\delta_t)$ should be first solved from Eq.\ (\ref{eq.e}) as follows:
\begin{equation}
    T = \dfrac{ \sum\nolimits_{i=0}^{8} g_i^{} - \tfrac{\delta_t}{2} c_v \bfu \cdot \nabla T }{ c_v - \tfrac{\delta_t}{2} (c_{v,s} - c_{v,f}) \partial f_s \big/ \partial t },
\end{equation}
and then $\epsilon (\bfx, t+\delta_t)$ and $q_c^{} (\bfx, t+\delta_t)$ can be immediately obtained.

\subsection{Particle dynamics}\label{sec.lbm.particle}
The governing equations of the solid particle motion can be written as \citep{Huang2012} 
\begin{subequations}\label{eq.particle}
    \begin{gather}
        \label{eq.particle.tran}
        M \dfrac{d \bfU_c}{d t} = \bfF_c ,
        \\
        \label{eq.particle.rota}
        \bfI_c \dfrac{d \bfOmega_c}{d t} + \bfOmega_c \times (\bfI_c \bfOmega_c) = \bfT_c ,
    \end{gather}
\end{subequations}
where $M$ is the mass of the solid particle, $\bfI_c$ is the inertia matrix of the solid particle with respect to its mass center, $\bfU_c$ is the velocity of the mass center, $\bfOmega_c$ is the angular velocity around the mass center, $\bfF_c$ is the total force acting on the solid particle, and $\bfT_c$ is the total torque acting on the solid particle with respect to its mass center. Based on the rigid body motion of the solid particle, the velocity of the solid fraction at lattice cell $\bfx$ and time $t$ can be immediately obtained
\begin{equation}\label{eq.us}
    \bfu_s (\bfx, t) = \bfU_c (t) + \bfOmega_c (t) \times [\bfx - \bfX_c (t)] ,
\end{equation}
where $\bfX_c$ denotes the mass center of the solid particle. Note that the solid particle is represented by a solid fraction field in the present volumetric LB method, meaning that the solid-fluid interface is smeared over one lattice cell rather than kept sharp. Under this scenario, the velocity of the solid fraction $\bfu_s$ at the solid/interface lattice cell $\bfx$ is always determined by the above equation, regardless of whether position $\bfx$ falls within the particle's sharp interface in reality. Such a simple strategy could be inaccurate from the viewpoint of a sharp interface but acceptable once the sharp interface is smeared over one lattice cell.

The key point to numerically solve Eq.\ (\ref{eq.particle}) is determining the hydrodynamic force and torque acting on the solid particle. On the basis of Eq.\ (\ref{eq.fi}), a specialized momentum exchange scheme can be formulated to calculate the hydrodynamic force and torque
\begin{subequations}
    \thinmuskip=1.5mu  \medmuskip=2.0mu  \thickmuskip=2.5mu
    \begin{gather}
        \bfF_\MES =  - \dfrac{\delta_x^D}{\delta_t} \sum\limits_{\bfx} \sum\limits_{i} (f_i - f_i^\ast) \bfe_i = - \dfrac{\delta_x^D}{\delta_t} \sum\limits_{\bfx} f_s \sum\limits_i \big[ f_i^\equ (\rho, \bfu_s) - f_i^\ast \big] \bfe_i ,
        \\
        \bfT_\MES = - \dfrac{\delta_x^D}{\delta_t} \sum\limits_{\bfx} (\bfx - \bfX_c) \times \sum\limits_{i} (f_i - f_i^\ast) \bfe_i = - \dfrac{\delta_x^D}{\delta_t} \sum\limits_{\bfx} (\bfx - \bfX_c) \times f_s \sum\limits_{i} \big[ f_i^\equ (\rho, \bfu_s) - f_i^\ast \big] \bfe_i ,
    \end{gather}
\end{subequations}
where $D$ is the dimension of space, and the summation with respect to $\bfx$ is performed for each particle. Since the present LB equation for velocity field is uniformly solved over the entire domain, the solid domain is \textit{numerically} filled with the fluid in the simulation. Therefore, the hydrodynamic force and torque calculated by the above momentum exchange scheme come from the fluids inside and outside the solid particle. In reality, the hydrodynamic force and torque acting on the solid particle should be merely from the outside fluid. As discussed in Sec.\ \ref{sec.lbm.velocity}, the nonslip velocity condition in the solid domain can be strictly satisfied by the present volumetric LB method, implying that the \textit{numerical fluid} inside the solid particle can be viewed as a \textit{pseudo rigid body} associated with the solid particle. The dynamics of this \textit{pseudo rigid body} are also described by Eq.\ (\ref{eq.particle}), and thus the force and torque acting on the solid particle from the inside fluid can be precisely expressed as 
\begin{subequations}\label{eq.inside.fluid}
    \begin{gather}
        \bfF_\ins = - M_\ins \dfrac{d \bfU_c}{d t} ,
        \\
        \bfT_\ins = - \bfI_{c,\ins} \dfrac{d \bfOmega_c}{d t} - \bfOmega_c \times (\bfI_{c,\ins} \bfOmega_c) ,
    \end{gather}
\end{subequations}
where $M_\ins = \rhof M / \rhos$ and $\bfI_{c,\ins} = \rhof \bfI_c / \rhos$ are the mass and inertia matrix of the inside fluid. The exact hydrodynamic force and torque acting on the solid particle can then be calculated by 
\begin{subequations}
    \begin{gather}
        \bfF_\outs = \bfF_\MES - \bfF_\ins ,
        \\
        \bfT_\outs = \bfT_\MES - \bfT_\ins .
    \end{gather}
\end{subequations}
In real applications, the total force $\bfF_c$ in Eq.\ (\ref{eq.particle.tran}) could consist of the hydrodynamic force $\bfF_\outs$ and the other body force like gravity or buoyancy, while the total torque $\bfT_c$ in Eq.\ (\ref{eq.particle.rota}) is equal to the hydrodynamic torque $\bfT_\outs$ as the other body force usually does not yield any torque with respect to the mass center. Note that once the solid particle collides with each other or with the wall, an additional repulsive force and torque should be introduced, and the collision model proposed by \citet{Glowinski2001} is adopted in this work. The reader is referred to previous works \citep{Feng2004, Glowinski2001, Pan2005} for more details.

In the simulations, $\bfF_\ins$ and $\bfT_\ins$ are explicitly calculated via Eq.\ (\ref{eq.inside.fluid}) with backward difference approximations for $d \bfU_c \big/ d t$ and $d \bfOmega_c \big/ d t$, and then the governing equations of the solid particle motion [i.e., Eq.\ (\ref{eq.particle})] are solved by the forward Euler method. As discussed by \citet{Suzuki2011} for the IB method, these explicit calculations of $\bfF_\ins$ and $\bfT_\ins$ won't introduce an additional restriction on the density ratio $\rhos / \rhof$ for numerical stability.

\section{Validations and discussions} \label{sec.val}
In this section, numerical tests are carried out to validate the present volumetric LB method for the PR DNS of thermal particulate flows with conjugate heat transfer. First, the transient conjugate heat transfer is considered to validate the present energy conservation equation [i.e., Eq.\ (\ref{eq.nse.e.final})] and the corresponding LB equation for temperature field. Then, the isothermal sedimentation of an elliptical particle, where many numerical results are available for comparison, is simulated to validate the LB equation for velocity field and the computation of the hydrodynamic force and torque. Afterward, the sedimentation of a cold particle with a fixed temperature is considered to demonstrate the capability of the present volumetric LB method for thermal particulate flows. Finally, the present method is applied to study particle sedimentations with conjugate heat transfer. In the simulations, the differences in thermophysical properties between the fluid and solid are characterized by the heat conductivity ratio $R_\lambda = \lambda_s / \lambda_f$ and the specific heat ratio $R_{c_v} = c_{v,s} / c_{v,f}$, where the specific heat of the fluid is fixed at $c_{v,f} = 1$. The relaxation parameters in $\bfS$ for the velocity field are set as $s_0 = s_j = 1$, $s_\varepsilon = s_e = 1.25$, $s_p = 1/\tau_f$, and $\big( s_p^{-1} - 0.5 \big) \big( s_q^{-1} - 0.5 \big) = 1/12$. The relaxation parameters in $\bfR$ for the temperature field are set as $\sigma_0 = 1$, $\sigma_\varepsilon = \sigma_e$, $\sigma_q = \sigma_j$, $\sigma_j = 1/\tau_g$, and $\big( \sigma_j^{-1} - 0.5 \big) \big( \sigma_e^{-1} - 0.5 \big) = 1/4$ unless otherwise stated. Here, the dimensionless relaxation times $\tau_f$ and $\tau_g$ are determined by the kinematic viscosity $\nu$ and heat conductivity $\lambda$, respectively.

\subsection{Transient conjugate heat transfer}
\begin{figure}[tbp]
    \centering
    \includegraphics[scale=1.0,draft=\figdraft]{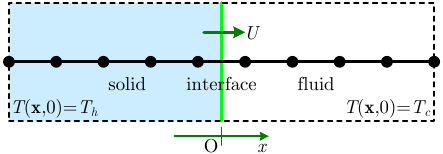}
    \caption[]{Schematic of the one-dimensional transient conjugate heat transfer with the coordinate, velocity, and initial condition shown. The solid-fluid interface is located at the halfway point between two adjacent lattice nodes at time $t=0$.}
    \label{fig.02}
\end{figure}

To validate the present energy conservation equation [i.e., Eq.\ (\ref{eq.nse.e.final})] and the corresponding LB equation for temperature field, the transient conjugate heat transfer is simulated. First, the one-dimensional case, where the analytical solution can be derived, is considered. As illustrated by Fig.\ \ref{fig.02}, the solid and fluid move right with constant velocity $(U, 0)^\Tr$. At time $t = 0$, the solid and fluid are in the domains $x<0$ and $x>0$, respectively, and the solid-fluid interface is located at $x = 0$. The temperature of the solid and fluid are fixed at $T_h$ and $T_c$ ($T_h > T_c$), respectively. As time goes on, the solid is cooled by the fluid, and the conjugate heat transfer condition is encountered at the solid-fluid interface. The analytical solution for this one-dimensional conjugate heat transfer problem is
\begin{equation}
    T (\bfx,t) = 
    \begin{dcases}
        T_h + (T_c - T_h) \dfrac{1}{ \sqrt{ R_\lambda R_{c_v} } + 1 } \mathrm{erfc} \bigg( \dfrac{X_i - x}{ 2 \sqrt{\alpha_s t} } \bigg) , & x < X_i \; (\mathrm{solid}) , \\
        T_c + (T_h - T_c) \dfrac{ \sqrt{ R_\lambda R_{c_v} } }{ \sqrt{ R_\lambda R_{c_v} } +1 } \mathrm{erfc} \bigg( \dfrac{x - X_i}{ 2 \sqrt{\alpha_f t} } \bigg) , & x > X_i \; (\mathrm{fluid}) , \\
    \end{dcases}
\end{equation}
where $\alpha_s = \lambda_s / (\rhof c_{v,s})$ and $\alpha_f = \lambda_f / (\rhof c_{v,f})$ are the exact thermal diffusivities of the solid and fluid, respectively, $X_i = U t$ is the location of the solid-fluid interface, and $\mathrm{erfc} (x) = \tfrac{2}{\sqrt{\pi}} \int\nolimits_x^{+\infty} e^{-\eta^2} d \eta$ is the complementary error function.

\begin{figure}[tbp]
    \centering
    \includegraphics[scale=1.0,draft=\figdraft]{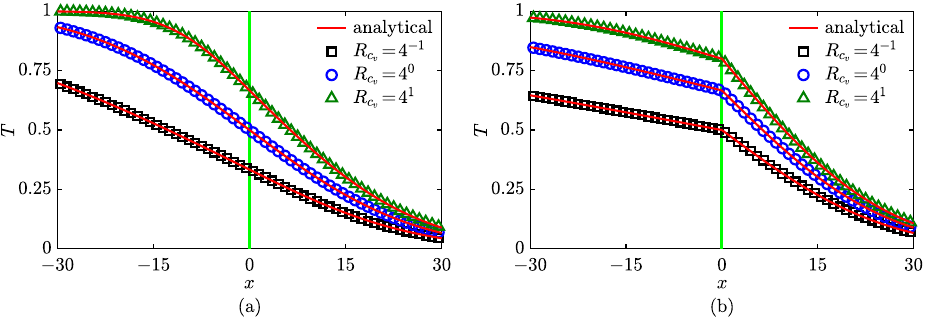}
    \caption[]{Temperature distribution near the solid-fluid interface at time $t = 2 \times 10^3$ for (a) $R_\lambda = 1$ and (b) $R_\lambda = 4$ in the stationary situation of the one-dimensional transient conjugate heat transfer. The symbols denote the numerical results and the solid lines denote the analytical solutions.}
    \label{fig.03}
\end{figure}

In the simulations, the lattice spacing and time step are set to $\delta_x = 1$ and $\delta_t = 1$, respectively. The initial temperature of the solid and fluid are fixed at $T_h = 1$ and $T_c = 0$, respectively. The density and heat conductivity of the fluid are set to $\rhof = 1$ and $\lambda_f = 0.1$, respectively. Note that only the LB equation for temperature field is involved here as the velocity field is specified in advance. The stationary situation with $U = 0$ is first considered, for which the source term $q_c^{}$ in Eq.\ (\ref{eq.nse.e.final}) vanishes because $\bfu = \mathbf{0}$ and $\partial f_s \big/ \partial t = 0$ over the entire domain. Figure \ref{fig.03} shows the temperature distribution near the solid-fluid interface at time $t = 2 \times 10^3$ and compares the numerical result with the analytical solution. The heat conductivity ratio is chosen as $R_\lambda = 1$ and $4$, and the specific heat ratio is chosen as $R_{c_v} = 4^{-1}$, $4^0$, and $4^1$. Good agreement between the numerical result and the analytical solution can be observed in Fig.\ \ref{fig.03}, which demonstrates that the conjugate heat transfer problem in the stationary situation can be correctly described by the present energy conservation equation [i.e., Eq.\ (\ref{eq.nse.e.final})] and accurately solved by the present LB equation for temperature field [i.e., Eq.\ (\ref{eq.lbe.e})].

\begin{figure}[tbp]
    \centering
    \includegraphics[scale=1.0,draft=\figdraft]{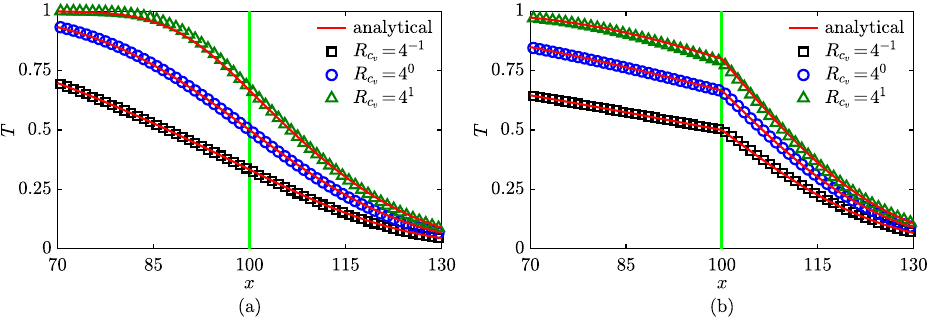}
    \caption[]{Temperature distribution near the solid-fluid interface at time $t = 2 \times 10^3$ for (a) $R_\lambda = 1$ and (b) $R_\lambda = 4$ in the moving situation of the one-dimensional transient conjugate heat transfer. The symbols denote the numerical results and the solid lines denote the analytical solutions.}
    \label{fig.04}
\end{figure}

To activate the source term $q_c^{}$ in Eq.\ (\ref{eq.nse.e.final}), the moving situation with $U = 0.05$ is then considered. As the solid-fluid interface moves, the solid fraction and the thermophysical properties vary with space and time near the solid-fluid interface. Figure \ref{fig.04} shows the comparisons of the temperature distribution near the solid-fluid interface between the numerical result and the analytical solution. The heat conductivity ratio is $R_\lambda = 1$ and $4$, and the specific heat ratio is $R_{c_v} = 4^{-1}$, $4^0$, and $4^1$. The chosen time is also $t = 2 \times 10^3$, and thus the solid-fluid interface moves to the position $X_i = 100$. It can be seen from Fig.\ \ref{fig.04} that the numerical result is in good agreement with the analytical solution, even at the lattice nodes just close to the solid-fluid interface. Such a good agreement demonstrates the applicability of the present energy conservation equation and the corresponding LB equation.

\begin{figure}[tbp]
    \centering
    \includegraphics[scale=1.0,draft=\figdraft]{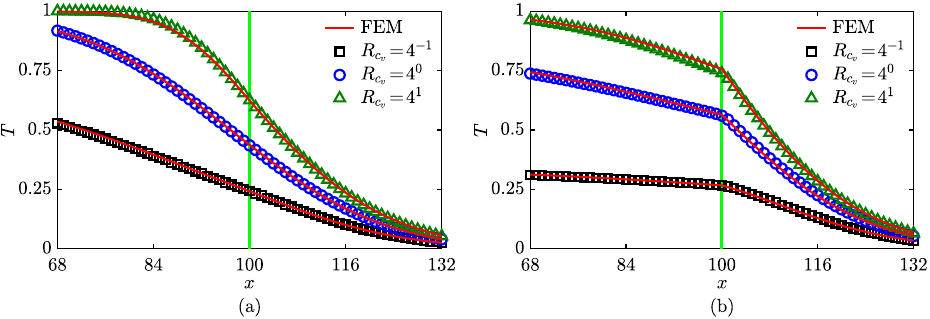}
    \caption[]{Radial temperature distribution near the solid-fluid interface at time $t = 2 \times 10^3$ for (a) $R_\lambda = 1$ and (b) $R_\lambda = 4$ in the moving situation of the two-dimensional transient conjugate heat transfer. The symbols denote the numerical results by the present LB method and the solid lines denote the benchmark solutions by the conventional FEM.}
    \label{fig.05}
\end{figure}

Then, the transient cooling of a solid circular cylinder by the fluid in the external infinite domain, as a two-dimensional case, is considered. The setups of this two-dimensional case remain the same as the above one-dimensional case, except for the geometric configuration. The origin of the reference frame is fixed at the initial center of the cylinder, and the diameter of the cylinder is set to $D = 64 \delta_x$. The solid-fluid system moves right with constant velocity $(0.05, 0)^\Tr$ to activate the source term $q_c^{}$ in Eq.\ (\ref{eq.nse.e.final}). Based on the Galilean invariance, it is immediately known that the radial temperature distribution for this moving situation should be the same as that for the stationary situation, which is obtained using the conventional FEM here and serves as the benchmark solution for validation. Figure \ref{fig.05} compares the numerical results by the present LB method with the benchmark solutions by the conventional FEM, where the time is chosen as $t = 2 \times 10^3$, and thus the rightmost position of the solid-fluid interface moves to $X_i = 100$. Good agreement, even at the lattice nodes just close to the solid-fluid interface, can also be observed in Fig.\ \ref{fig.05}, which reconfirms the applicability of the present energy conservation equation and the corresponding LB equation.

\subsection{Isothermal sedimentation of an elliptical particle}\label{sec.sed.iso}
\begin{figure}[tbp]
    \centering
    \includegraphics[scale=1.0,draft=\figdraft]{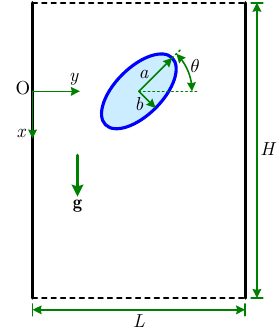}
    \caption[]{Schematic of the isothermal sedimentation of an elliptical particle in a long channel with the coordinate, gravity acceleration, and geometric parameters shown. The initial position of the mass center of the elliptical particle is set to $(0, 0.5L)^\Tr$.}
    \label{fig.06}
\end{figure}

To validate the LB equation for velocity field and the computation of the hydrodynamic force and torque acting on the solid particle, the isothermal sedimentation of an elliptical particle in a long channel, where many numerical results are available for comparison, is considered. Compared with the circular particle, simulating the sedimentation of the elliptical particle is relatively more sensitive to the computational accuracy of the hydrodynamic force and torque. As illustrated by Fig.\ \ref{fig.06}, the major and minor axes of the elliptical particle are $a = 0.05 \,\cm$ and $b = 0.025 \,\cm$, respectively, and the width of the channel is $L = 0.4 \,\cm$. The densities of the solid and fluid are $\rhos = 1.1 \,\g / \cm^3$ and $\rhof = 1.0 \,\g / \cm^3$, respectively. The kinematic viscosity of the fluid is $\nu = 0.01 \,\cm^2 / \s$, and the gravity acceleration is $|\bfg| = 980 \,\cm / \s^2$. Since the solid density is larger than the fluid density, the elliptical particle settles down due to gravity.

\begin{figure}[tbp]
    \centering
    \includegraphics[scale=1.0,draft=\figdraft]{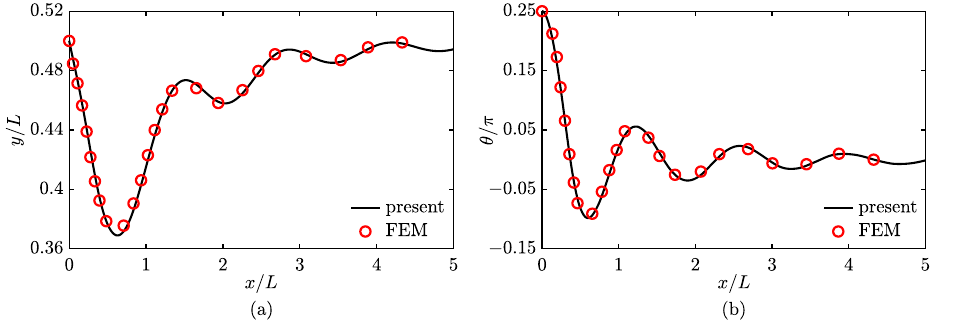}
    \caption[]{Comparisons of (a) the particle trajectory ($y/L$ versus $x/L$) and (b) the particle orientation ($\theta / \pi$ versus $x/L$) between the present results and the previous FEM results by \citet{Xia2009}.}
    \label{fig.07}
\end{figure}

In the simulations, the channel width is divided into $104$ lattice cells, implying that the lattice spacing is $\delta_x = 1/260 \,\cm$. To mimic the long channel, the height of the computational domain is chosen as $H = 30L$, and the no-flow condition with $\bfu = \mathbf{0}$ and the fully developed condition with $\partial \bfu / \partial x = \mathbf{0}$ are imposed at the upper and lower sides of the computational domain, respectively. The dimensionless relaxation time for the density distribution function is set to $\tau_f = 0.6$, and then the time step can be determined via the relation $\nu = c_s^2 \delta_t \big( s_p^{-1} - 0.5 \big)$. At time $t = 0$, the velocity field is initialized as $\bfu = \mathbf{0}$ over the entire domain, and the elliptical particle is placed in the middle of the channel and $3L$ away from the upper side of the computational domain. The orientation of the particle, i.e., the angle between its major axis and the horizontal (the $y$ coordinate shown in Fig.\ \ref{fig.06}), is initialized as $\theta = \pi / 4$. Then, the elliptical particle starts settling down in the channel. Note that only the LB equation for velocity field is involved here because the sedimentation process is isothermal. In practice, a net body force $(\rhos - \rhof) \bfg$ can be applied to the solid particle, and then the fluid is free of any body force. The particle trajectory ($y/L$ versus $x/L$) and orientation ($\theta / \pi$ versus $x/L$) in the sedimentation process are plotted in Fig.\ \ref{fig.07} and compared with previous results by the conventional FEM. Here, $(x, y)^\Tr$ denotes the location of the mass center of the elliptical particle. Good agreement between the present results with the previous FEM results can be observed, demonstrating the applicability of the present LB equation for velocity field and the computational accuracy of the hydrodynamic force and torque. It can also be seen from Fig.\ \ref{fig.07} that, at the initial stage, the elliptical particle swings towards the left wall, accompanied by a counterclockwise rotation around its mass center. As time goes on, the elliptical particle tends to settle down along the channel's centerline, with its orientation tending to zero.

\begin{figure}[tbp]
    \centering
    \includegraphics[scale=1.0,draft=\figdraft]{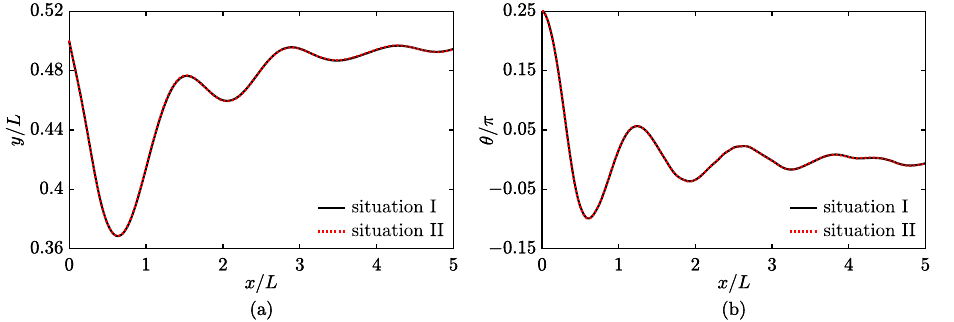}
    \caption[]{Comparisons of (a) the particle trajectory ($y/L$ versus $x/L$) and (b) the particle orientation ($\theta / \pi$ versus $x/L$) between the situations when the fluid is free of any body force (situation I) and when the fluid is subjected to gravity (situation II).}
    \label{fig.08}
\end{figure}

Before proceeding further, it is necessary to validate that the present LB equation for velocity field and the computation of the hydrodynamic force and torque apply to the situation when the fluid itself is subjected to body force. For this purpose, the above simulation is repeated by directly applying the gravity forces $\rhos \bfg$ and $\rhof \bfg$ to the solid and fluid, respectively. It is worth pointing out that a pressure gradient will be induced inside the fluid by the gravity force, which should be carefully considered in the initialization and boundary condition treatment of the simulation. Furthermore, this pressure gradient will lead to a density gradient as the pressure is determined by $p = \rho c_s^2$ in the LB method. Thus, to well approximate the incompressible condition of the fluid, the dimensionless relaxation time $\tau_f$ is changed from $0.6$ to $0.515$ to increase the lattice sound speed $c_s$ and then suppress the density variation caused by the pressure gradient. Figure \ref{fig.08} shows the particle trajectory ($y/L$ versus $x/L$) and orientation ($\theta / \pi$ versus $x/L$) in the sedimentation process and compares the results with these for the situation when the fluid is free of any body force, which is simulated again with $\tau_f = 0.515$ for the sake of comparison. As expected, there is no distinguishable difference between the numerical results for the situations when the fluid is free of any body force and when the fluid is subjected to gravity. Therefore, the present LB equation for velocity field and the computation of the hydrodynamic force and torque apply when the fluid is subjected to body force and thus is capable of handling thermal particulate flows where buoyancy should be considered.

\subsection{Sedimentation of a cold particle with fixed temperature}\label{sec.sed.ft}
\begin{figure}[tbp]
    \centering
    \includegraphics[scale=1.0,draft=\figdraft]{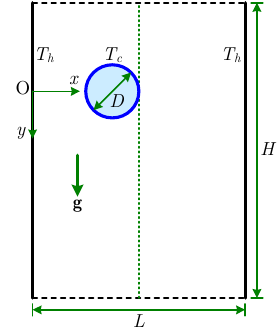}
    \caption[]{Schematic of the sedimentation of a cold particle with fixed temperature in a long channel, where the coordinate, gravity acceleration, and geometric parameters are given. The initial position of the mass center of the circular particle is set to $(1.5D, 0)^\Tr$.}
    \label{fig.09}
\end{figure}

To demonstrate the capability of the present volumetric LB method for thermal particulate flows, the sedimentation of a cold particle with fixed temperature in a long channel is simulated in this section. Since the temperature of the solid particle is assumed to be constant, neither the heat transfer inside the solid particle nor the conjugate heat transfer condition at the solid-fluid interface is involved. Thus, this sedimentation process is only affected by the density but not the other thermophysical properties of the solid particle. As shown in Fig.\ \ref{fig.09}, the width of the long channel is $L = 4D$ with $D$ the diameter of the circular particle. At time $t = 0$, the temperatures of the solid particle and fluid are uniformly set to $T_c$ and $T_h$ ($T_c < T_h$), respectively, and the particle center deviates from the channel's centerline with a distance of $D/2$. In the sedimentation process, the temperatures of the solid particle and the channel walls are kept at $T_c$ and $T_h$, respectively. A net body force $(\rhos - \rhof) \bfg$ is applied to the solid particle, and the buoyancy force $-\rhof \bfg \beta (T - T_\refe)$ is applied to the fluid, where the reference temperature $T_\refe$ is set to $T_h$. Four dimensionless parameters:\ the density ratio $\rhor$, the Reynolds number $\Reyn$, the Prandtl number $\Prat$, and the Grashof number $\Gras$, can be introduced to characterize this sedimentation process, which are defined as
\begin{equation}\label{eq.nodim.paras}
    \rhor = \dfrac{\rhos}{\rhof} ,\quad \Reyn = \dfrac{U_\refe D}{\nu} ,\quad \Prat = \dfrac{\nu}{\alpha} ,\quad \Gras = \dfrac{ |\bfg| \beta (T_h - T_c) D^3  }{\nu^2}.
\end{equation} 
Here, $U_\refe = \sqrt{ (\rhor-1) |\bfg| \pi D /2}$ is a reference velocity with $\bfg$ the gravity acceleration, and $\nu$, $\alpha$, and $\beta$ are the kinematic viscosity, thermal diffusivity, and volume expansivity of the fluid, respectively.

Following the work by \citet{Yu2006}, the density ratio, Reynolds number, and Prandtl number are set to $\rhor = 1.00232$, $\Reyn = 40.5$, and $\Prat = 0.7$, respectively, and the Grashof number is chosen as $\Gras = 100$, $564$, $1000$, $2000$, $2500$, and $4500$, respectively, which correspond to five different regimes for the particle motion defined by \citet{Gan2003}. The basic simulation parameters are set to $D = 1$, $\delta_x = D/20$, $\delta_t = 0.004$, $T_c = 0$, $T_h = 1$, $\rhof = 1$, and $c_{v,f} = 1$. The dimensionless relaxation time for the density distribution function is chosen as $\tau_f = 0.53$. Since the heat conductivity and specific heat of the solid do not take effect in the sedimentation process, the heat conductivity ratio $R_\lambda$ and the specific heat ratio $R_{c_v}$ are fixed at $1$ in the simulations.

\begin{figure}[tbp]
    \centering
    \includegraphics[scale=1.0,draft=\figdraft]{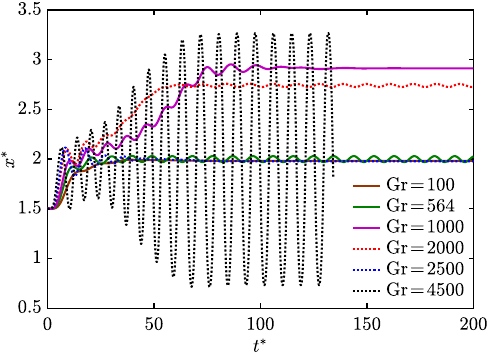}
    \caption[]{Time evolution of the horizontal position of the particle center in the sedimentation process, where the horizontal position and time are normalized as $x^\ast = x / D$ and $t^\ast = U_\refe t / D$, respectively.}
    \label{fig.10}
\end{figure}

Figure \ref{fig.10} shows the time evolutions of the horizontal position of the particle center in the sedimentation process, where the horizontal position and time are normalized as $x^\ast = x/D$ and $t^\ast = U_\refe t / D$, respectively. It can be seen from Fig.\ \ref{fig.10} that the time evolution of $x^\ast$ at the fully developed stage is strongly dependent on the Grashof number $\Gras$. All five regimes [including the steady settling down along the channel's centerline ($\Gras = 100$), the periodic oscillating around the channel's centerline ($\Gras = 564$), the settling down close to one wall with/without oscillation ($\Gras = 1000$ and $2000$), the settling down along the channel's centerline without oscillation ($\Gras = 2500$), and the periodic oscillating around the channel's centerline with a large amplitude ($\Gras = 4500$)] are successfully reproduced in the present simulations, and the curves shown in Fig.\ \ref{fig.10} are in good agreement with previous works (see Fig.\ 7 in \citet{Yu2006} and Fig.\ 8 in \citet{Kang2011.B}). For quantitative comparison, the equilibrium horizontal position (i.e., the time-averaged $x^\ast$ at the fully developed stage) for $\Gras = 1000$ and $2000$, as well as the oscillation amplitude of $x^\ast$ at the fully developed stage for $\Gras = 4500$, are calculated and listed in Table \ref{tab.04}. It can be seen from Table \ref{tab.04} that the present results agree well with previous results \citep{Yu2006, Kang2011.B, Feng2009, Mozafari2018}, which demonstrates that the present volumetric LB method can accurately handle thermal particulate flows when the temperature of the solid particle is assumed to be constant.

\begin{table}[tbp]
    \centering
    \caption{Comparisons of the equilibrium horizontal position for $\Gras = 1000$ and $2000$ and the oscillation amplitude of the horizontal position at the fully developed stage for $\Gras = 4500$ between the present and previous results, where the horizontal position is normalized as $x^\ast = x / D$. }
    \label{tab.04}
    \setlength{\tabcolsep}{44.5pt}
    \begin{tabular}{lccc}
        \hline\hline
        $\Gras$                 & $1000$ & $2000$ & $4500$ \\
        \hline
        Present                 & $2.91$ & $2.74$ & $1.27$ \\
        \citet{Yu2006}          & $2.89$ & $2.74$ & $1.32$ \\
        \citet{Kang2011.B}      & $2.91$ & $2.74$ & $1.32$ \\
        \citet{Feng2009}        & $2.90$ & $2.73$ & $1.35$ \\
        \citet{Mozafari2018}    & $2.76$ & $2.62$ & $1.30$ \\
        \hline\hline
    \end{tabular}
\end{table}

\subsection{Sedimentation of a cold particle with conjugate heat transfer}\label{sec.sed.cht}
To show the versatility of the present volumetric LB method, the sedimentation of a cold particle in a long channel, where the temperature field inside the solid particle is fully resolved and thus the conjugate heat transfer condition at the solid-fluid interface is involved, is simulated in this section. The schematic of this problem is also given in Fig.\ \ref{fig.09}, where the mass center of the circular particle is moved to the channel's centerline at time $t=0$. The four dimensionless parameters, also defined in Eq.\ (\ref{eq.nodim.paras}), are fixed at $\rhor = 1.00232$, $\Reyn =10.5$, $\Prat = 0.7$, and $\Gras = 1000$, respectively. The other simulation parameters are chosen the same as those in Sec.\ \ref{sec.sed.ft}. Under such configurations, this sedimentation process with conjugate heat transfer is now characterized by the heat conductivity ratio $R_\lambda = \lambda_s / \lambda_f$ and the specific heat ratio $R_{c_v} = c_{v,s} / c_{v,f}$.

\begin{figure}[tbp]
    \centering
    \includegraphics[scale=1.0,draft=\figdraft]{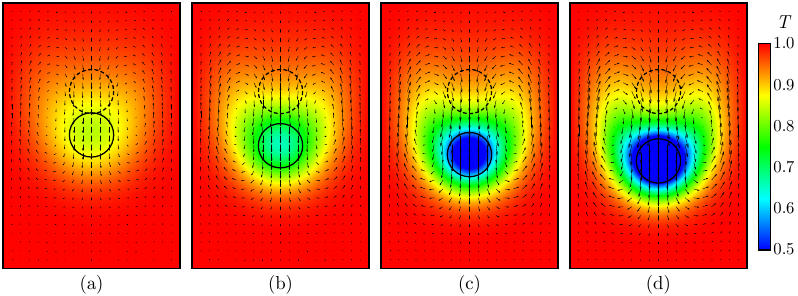}
    \caption[]{Temperature field, together with the velocity vectors, in the sedimentation process at the normalized time $t^\ast = 2.625$ for $R_\lambda = 1$ and (a) $R_{c_v} = 1$, (b) $R_{c_v} = 2$, (c) $R_{c_v} = 4$, and (d) $R_{c_v} = 8$. The dashed and solid circles denoted the initial and current positions of the solid particle, respectively.}
    \label{fig.11}
\end{figure}

\begin{figure}[tbp]
    \centering
    \includegraphics[scale=1.0,draft=\figdraft]{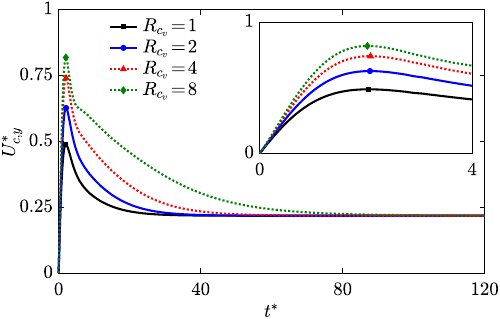}
    \caption[]{Time evolution of the normalized vertical velocity of the particle center in the sedimentation process for $R_\lambda = 1$ and $R_{c_v} = 1$, $2$, $4$, and $8$. The symbol denotes the maximum vertical velocity achieved in the sedimentation process.}
    \label{fig.12}
\end{figure}

First, the heat conductivity ratio is fixed at $R_\lambda = 1$, and the specific heat ratio is set to $R_{c_v} = 1$, $2$, $4$, and $8$, respectively. Figure \ref{fig.11} shows the temperature field in the sedimentation process at the normalized time $t^\ast = U_\refe t / D = 2.625$, where the velocity vectors and the initial and current positions of the particle are also depicted. It can be seen from Fig.\ \ref{fig.11} that the cold particle is heated at a smaller rate as the specific heat ratio $R_{c_v}$ increases. To be quantitative, the minimal temperatures inside the particle at time $t^\ast = 2.625$ are $0.841$, $0.656$, $0.399$, and $0.162$ for $R_{c_v} = 1$, $2$, $4$, and $8$, respectively. Meanwhile, a pair of vortices, rotating in different directions, are induced in the wake of the particle. At the early stage of the sedimentation process, the fluid near the particle becomes cold. It thus sinks due to the buoyancy force $-\rhof \bfg \beta (T - T_\refe)$, which accelerates the sedimentation of the particle. Therefore, the particle with a relatively large $R_{c_v}$ settles down faster, as observed in Fig.\ \ref{fig.11}. To better understand the particle sedimentation with conjugate heat transfer, the time evolution of the vertical velocity of the particle center in the sedimentation process is plotted in Fig.\ \ref{fig.12}. Here, the vertical velocity is normalized as $U_{c,y}^\ast = U_{c,y} / U_\refe$. It can be seen from Fig.\ \ref{fig.12} that the time evolution can be sequentially divided into three stages: the accelerating stage, the decelerating stage, and the equilibrium stage. At the accelerating stage, the normalized vertical velocity $U_{c,y}^\ast$ rapidly increases due to the cooperative effects of the net body force $(\rhos - \rhof) \bfg$ exerted on the particle and the buoyancy force $-\rhof \bfg \beta (T - T_\refe)$ exerted on its adjacent fluid, which becomes cold due to the conjugate heat transfer there. As time goes on, the fluid gradually heats the cold particle, and the temperature of the adjacent fluid increases, decreasing the buoyancy force exerted on the adjacent fluid. Thus, the particle is slowed down by the drag force from the fluid, which corresponds to the decelerating stage. At the equilibrium stage, the cold particle is sufficiently heated by the fluid, and the temperature over the entire domain tends to $T_h$, implying that the sedimentation process with conjugate heat transfer degenerates into an isothermal situation. It is also interesting to note from Fig.\ \ref{fig.12} that the duration time of the accelerating stage is almost independent of the specific heat ratio $R_{c_v}$, while the maximum $U_{c,y}^\ast$ significantly increases as $R_{c_v}$ increases.

\begin{figure}[tbp]
    \centering
    \includegraphics[scale=1.0,draft=\figdraft]{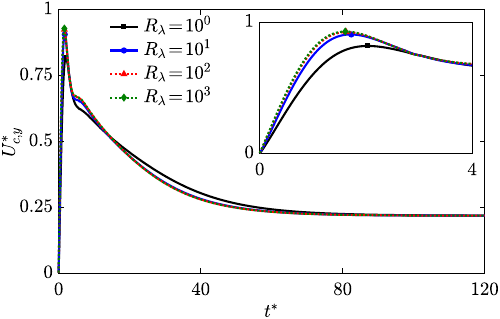}
    \caption[]{Time evolution of the normalized vertical velocity of the particle center in the sedimentation process for $R_{c_v} = 8$ and $R_\lambda = 10^0$, $10^1$, $10^2$, and $10^3$. The symbol denotes the maximum vertical velocity achieved in the sedimentation process.}
    \label{fig.13}
\end{figure}

Then, the influence of the heat conductivity ratio $R_\lambda$ on the sedimentation process is investigated by fixing $R_{c_v} = 8$ and varying $R_\lambda$ from $10^0$ to $10^3$. Figure \ref{fig.13} shows the time evolution of the normalized vertical velocity $U_{c,y}^\ast$ in the sedimentation process. It can be seen from Fig.\ \ref{fig.13} that, as $R_\lambda$ varies from $10^0$ to $10^3$, the duration time of the accelerating stage monotonously decreases from $2.032$ to $1.614$, while the maximum $U_{c,y}^\ast$ monotonously increases from $0.818$ to $0.928$. Furthermore, these variation trends are relatively significant when $R_\lambda$ is small (such as $R_\lambda \leq 10^1$) and almost visually indistinguishable when $R_\lambda$ is large enough (such as $R_\lambda \geq 10^2$). These phenomena observed in Fig.\ \ref{fig.13} can be explained as follows:\ As the heat conductivity ratio $R_\lambda$ increases, the heat conduction inside the solid particle becomes faster, implying that the conjugate heat transfer between the particle and fluid is enhanced. Thus, the influence of the buoyancy force $-\rhof \bfg \beta (T - T_\refe)$, exerted on the adjacent fluid, on the particle sedimentation is concentrated in time, which leads to the increase of the maximum $U_{c,y}^\ast$ and the decrease of the duration time of the accelerating stage. When $R_\lambda$ is large enough, the conjugate heat transfer is dominated by the heat convection inside the fluid. Thus, the influence of increasing $R_\lambda$ on the sedimentation process becomes relatively weak.

\subsection{Sedimentation of 2048 cold particles with conjugate heat transfer}
As a further application of the present volumetric LB method to dense particulate flows, the sedimentation of 2048 cold particles in a square cavity is simulated in this section. The temperature fields over the solid and fluid domains are fully resolved, and the conjugate heat transfer between the particle and fluid is involved. All the four walls of the cavity are adiabatic. The lattice spacing is fixed at $\delta_x = 0.01$. The diameter of the circular particle is chosen as $D = 24 \delta_x$, and the side length of the square cavity is set to $L = 1700 \delta_x$. At time $t = 0$, the 2048 particles are arranged in $32$ rows and $64$ columns, and both the horizontal and vertical gaps between two adjacent particles are fixed at $2 \delta_x$. The initial gap between the left wall and the leftmost particle is set to $19 \delta_x$, and the initial gap between the upper wall and the uppermost particle is set to $18 \delta_x$. The temperatures of the particles and fluid are initialized as $T_c = 0$ and $T_h = 1$, respectively. In the simulation, a net body force $(\rhos - \rhof) \bfg$ is applied to the particle, and the buoyancy force $-\rhof \bfg \beta (T - T_\refe)$, with the reference temperature $T_\refe = T_h$, is applied to the fluid. The four dimensionless parameters, also defined in Eq.\ (\ref{eq.nodim.paras}), are chosen as $\rhor = 1.5$, $\Reyn = 40.5$, $\Prat = 0.71$, and $\Gras = 100$, respectively. The heat conductivity and specific heat ratios are fixed at $R_\lambda = 1$ and $R_{c_v} = 2$, respectively. The density, specific heat, and kinematic viscosity of the fluid are chosen as $\rhof = 1$, $c_{v,f} = 1$, and $\nu = 0.01$, respectively. The dimensionless relaxation time for the density distribution function is set to $\tau_f = 0.55$, and the relaxation parameters in $\bfR$ for the temperature field are chosen as $\sigma_0 = \sigma_\varepsilon = \sigma_e = 1$ and $\sigma_q = \sigma_j = 1/\tau_g$ to achieve better numerical stability in the simulation.

\begin{figure}[tbp]
    \centering
    \includegraphics[scale=1.0,draft=\figdraft]{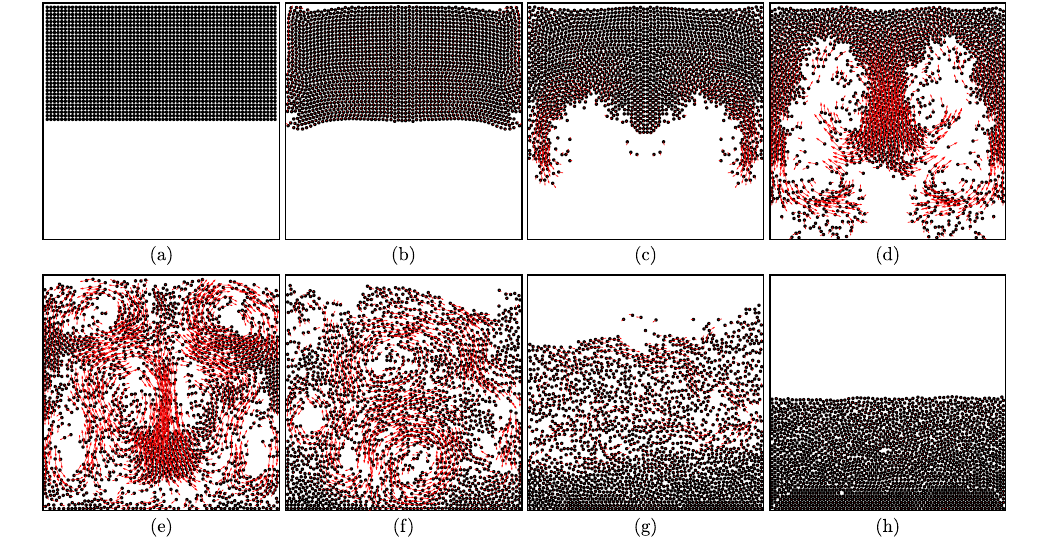}
    \caption[]{Snapshots of the sedimentation of 2048 cold particles in a square cavity at the normalized time (a) $t^\ast = 0.00$, (b) $t^\ast = 21.04$, (c) $t^\ast = 42.08$, (d) $t^\ast = 63.12$, (e) $t^\ast = 84.16$, (f) $t^\ast = 126.24$, (g) $t^\ast = 210.41$, and (h) $t^\ast = 420.81$, where the arrow denotes the velocity vector of the mass center of the particle.}
    \label{fig.14}
\end{figure}

\begin{figure}[tbp]
    \centering
    \includegraphics[scale=1.0,draft=\figdraft]{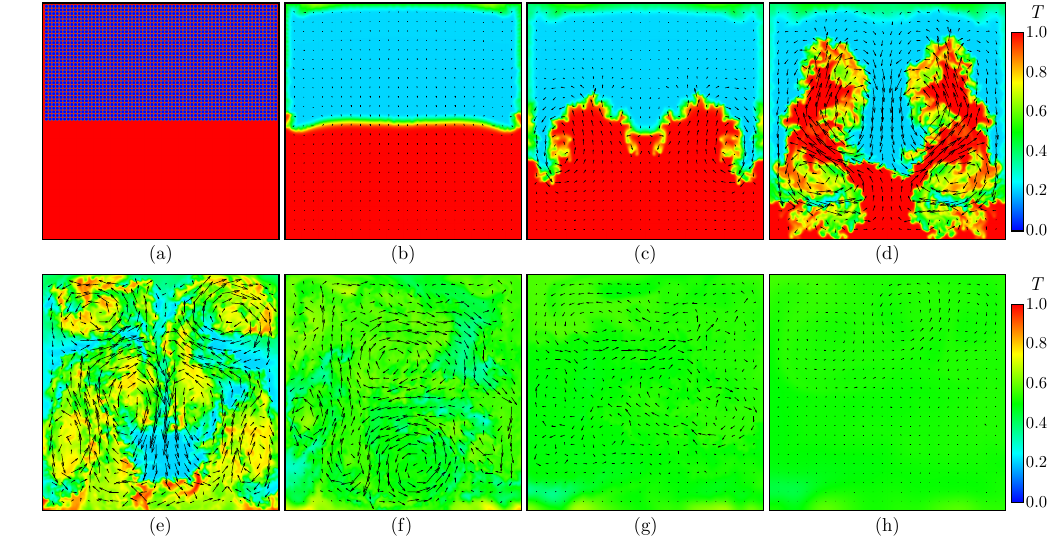}
    \caption[]{Snapshots of the temperature and velocity fields in the sedimentation of 2048 cold particles in a square cavity at the normalized time (a) $t^\ast = 0.00$, (b) $t^\ast = 21.04$, (c) $t^\ast = 42.08$, (d) $t^\ast = 63.12$, (e) $t^\ast = 84.16$, (f) $t^\ast = 126.24$, (g) $t^\ast = 210.41$, and (h) $t^\ast = 420.81$.}
    \label{fig.15}
\end{figure}

The snapshots of the sedimentation of 2048 cold particles are shown in Fig.\ \ref{fig.14}, where the time is normalized as $t^\ast = U_\refe t / D$ with the reference velocity $U_\refe = \sqrt{(\rhor - 1) |\bfg| \pi D /2}$. The snapshots of the temperature and velocity fields at the same moments as Fig.\ \ref{fig.14} are shown in Fig.\ \ref{fig.15}. It can be seen from Fig.\ \ref{fig.14}(b) that the particles closest to the left and right walls move upwards instead of settling down at $t^\ast = 21.04$, which is caused by the rising fluid there as the initial gaps between the vertical walls and particles are relatively large. Such an unexpected phenomenon is also observed in previous works for isothermal sedimentation \citep{Nie2010, Feng2009}. At the same moment as shown in Fig.\ \ref{fig.15}(b), the fluid in the upper part of the cavity (i.e., within the interparticle gaps) is sufficiently cooled down, while the fluid in the lower part of the cavity remains hot. Due to the initial temperature difference between the particle and fluid, the conjugate heat transfer between the particle and fluid is very strong at the initial stage of the sedimentation process, and the temperature field in the upper part of the cavity becomes relatively homogeneous at $t^\ast = 21.04$. As time goes on, the particles near the left and right walls, as well as the cavity's centerline, settle down more quickly than the other particles, which can be observed in Fig.\ \ref{fig.14}(c) and (d) for $t^\ast = 42.08$ and $63.12$, respectively. Corresponding to this sedimentation, the low-temperature region in the upper part of the cavity penetrates into the high-temperature region in the lower part of the cavity, as shown in Fig.\ \ref{fig.15}(c) and (d). Moreover, two large vortexes are induced near the left and right walls at $t^\ast = 42.08$, each of which splits into two vortexes, with one close to the bottom of the cavity and the other moving upward, at $t^\ast = 63.12$. Here, it is worth pointing out that the sedimentation process is fully symmetric concerning the cavity's centerline when $t^\ast \leq 42.08$ and becomes slightly left-right asymmetric at $t^\ast = 63.12$ when the particles start reaching the lower wall of the cavity. After that, the sedimentation process is fully left-right asymmetric. The snapshots at $t^\ast = 42.08$ and $63.12$ indicate that the particulate Rayleigh-B\'{e}nard convection is successfully captured in the present PR DNS. At time $t^\ast = 84.16$, the particles are distributed almost over the whole cavity [see Fig.\ \ref{fig.14}(e)], and the fluid flow is quite chaotic and relatively intense [see Fig.\ \ref{fig.15}(e)]. Thus, the influence of gravity on particle motion is relatively weak, and the particles are mainly dragged by the fluid. When $t^\ast \geq 126.24$, the particles start packing at the bottom of the cavity, and the temperature field tends to be uniform over the entire cavity. It can be seen from Fig.\ \ref{fig.14}(f), (g), and (h) that the packing process is relatively slow, and the particles at the lower layer pack more closely. Moreover, this packing process also demonstrates that the present volumetric LB method is quite robust when the particles are closely packed, which is expected as all the computations in the present method are locally performed except the linear streaming process of the LB equation. As shown by Fig.\ \ref{fig.15}(h), the temperature field is visually uniform over the entire cavity at $t^\ast = 420.81$. The corresponding space-average temperature is about $0.5199$, close to the analytical equilibrium temperature $T_\equ = 0.5145$ calculated by the energy balance relation. This agreement between the space-average temperature at $t^\ast = 420.81$ and the analytical equilibrium temperature indicates the validity of the present volumetric LB method for the PR DNS of thermal particulate flows with conjugate heat transfer.

\section{Conclusions} \label{sec.con}
This work develops a volumetric LB method as a single-domain approach for the particle-resolved direct numerical simulation of thermal particulate flows with conjugate heat transfer. The fluid flow and heat transfer around every particle are fully resolved, and the complete information on thermal particulate flows can be provided. The nonslip velocity condition in the solid domain is strictly satisfied, and the hydrodynamic force and torque acting on the solid particle are accurately calculated. The differences in thermophysical properties between the solid and fluid are correctly handled, and the numerical fidelity is well preserved even very close to the solid-fluid interface. Except for the linear streaming process of the LB equation, all the computations in the present method are locally performed, ensuring its robustness for closely packed particles and suitability for massively parallel computing.

Numerical simulations of transient conjugate heat transfer, isothermal sedimentation of an elliptical particle, and sedimentation of a cold particle with fixed temperature are first carried out. The present results agree well with the analytical/benchmark solutions and previous results, which validate the present volumetric LB method in various aspects. Then, the sedimentation of a cold particle with conjugate heat transfer in a long channel is investigated. It is found that this sedimentation process can be divided into three stages: the accelerating stage, the decelerating stage, and the equilibrium stage. As the solid-to-fluid specific heat ratio increases, the maximum sedimentation velocity increases significantly while the accelerating stage lasts almost the same time. As the solid-to-fluid heat conductivity ratio increases, the maximum sedimentation velocity slightly increases while the accelerating stage shortens. Finally, the sedimentation of 2048 cold particles with conjugate heat transfer in a square cavity is simulated. At the early stage, the conjugate heat transfer between the particle and fluid is strong, and the particles near the left and right walls, as well as the cavity’s centerline, settle down faster, resulting in the pattern of particulate Rayleigh-Bénard convection. At the end stage, the particles closely pack at the bottom of the cavity, and this packing process is relatively slow compared with the early sedimentation process.

\begin{acknowledgments}
    This work was supported by the National Natural Science Foundation of China through Grants No.\ 52376086 and No.\ 52006244.
\end{acknowledgments}

\appendix
\section{Theoretical analysis} \label{app.analysis}
\begin{figure}[tbp]
    \centering
    \includegraphics[scale=1.0,draft=\figdraft]{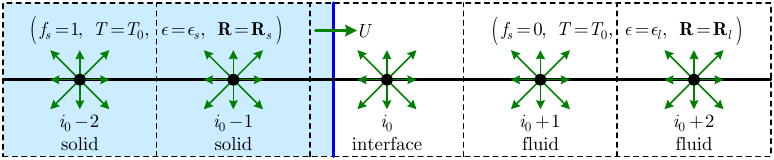}
    \caption[]{Schematic of the one-dimensional isothermal system moving right with a constant velocity $(U, 0)^\Tr$. The solid and fluid are in the left and right semi-infinite domains. The circular point denotes the lattice node, the dashed line denotes the lattice grid with the forming dashed box the lattice cell, and the arrow starting from the lattice node denotes the lattice discrete velocity. }
    \label{fig.16}
\end{figure}

To determine the parameters in the LB equation for temperature field and the computation of the time derivative $\partial f_s \big/ \partial t$ in the source term, a one-dimensional isothermal system is theoretically analyzed here. As illustrated by Fig.\ \ref{fig.16}, the solid and fluid are in the left and right semi-infinite domains, respectively, and the thermophysical properties of the solid and fluid are different (i.e., $c_{v,s} \neq c_{v,f}$ and $\lambda_s \neq \lambda_f$), implying that $\epsilon_s \neq \epsilon_f$ and $\bfR_s \neq \bfR_f$. The solid-fluid system is initially isothermal and moves right with a constant velocity $(U, 0)^\Tr$. Based on the Galilean invariance, it is immediately known that the solid-fluid system should analytically stay at the initial temperature $T_0$. To simplify the notations, the position $x$ and time $t$ are denoted by the superscripts ``$i$'' and ``$n$'', respectively, and thus $i \pm 1$ and $n \pm 1$ represent $x \pm \delta_x$ and $t \pm \delta_t$, respectively. For a one-dimensional problem in $x$ direction, the summation of the internal energy distribution functions $g_0^{}$, $g_2^{}$, and $g_4^{}$ can be calculated from the LB equation [i.e., Eq.\ (\ref{eq.lbe.e})] as
\begin{subequations}
    \begin{equation} \narrow{1.5mu} \label{eq.g024}
        g_{0+2+4}^{i,n+1} = \left( 1 - \sigma_e^{i,n} \right) g_{0+2+4}^{i,n} + \sigma_e^{i,n} \left( \epsilon^{i,n} - \dfrac{4+\alpha_1}{6} c_{v,\refe} T^{i,n} \right) + \dfrac{2-\sigma_e^{i,n}}{2} \dfrac{2-\beta_1}{6} \delta_t q_c^{i,n},
    \end{equation}
where the subscript ``$0+2+4$'' implies $g_0^{} + g_2^{} + g_4^{}$. Similarly, the summation of $g_1^{}$, $g_5^{}$, and $g_8^{}$, denoted by $g_{1+5+8}^{}$, can be calculated as 
\begin{equation} \narrow{1.5mu} \label{eq.g158}
    \begin{split}
        g_{1+5+8}^{i,n+1} = {} & \left( 1 - \sigma_j^{i-1,n} \right) g_{1+5+8}^{i-1,n} - \dfrac{ \sigma_j^{i-1,n} - \sigma_e^{i-1,n} }{2} g_{0+2+4}^{i-1,n} \\ {} & + \dfrac{ \sigma_j^{i-1,n} - \sigma_e^{i-1,n} }{2} \epsilon^{i-1,n} + \dfrac{ \sigma_e^{i-1,n} }{2} \dfrac{ 4+\alpha_1 }{6} c_{v,\refe} T^{i-1,n} + \left( \dfrac{ 2-\sigma_j^{i-1,n} }{4} - \dfrac{ 2-\sigma_e^{i-1,n} }{2} \dfrac{ 2-\beta_1 }{12} \right) \delta_t q_c^{i-1,n} ,  
    \end{split}
\end{equation}
and the summation of $g_3^{}$, $g_6^{}$, and $g_7^{}$, denoted by $g_{3+6+7}^{}$, can be calculated as 
\begin{equation} \narrow{1.5mu} \label{eq.g367}
    \begin{split}
        g_{3+6+7}^{i,n+1} = {} & \left( 1 - \sigma_j^{i+1,n} \right) g_{3+6+7}^{i+1,n} - \dfrac{ \sigma_j^{i+1,n} - \sigma_e^{i+1,n} }{2} g_{0+2+4}^{i+1,n} \\ {} & + \dfrac{\sigma_j^{i+1,n} - \sigma_e^{i+1,n} }{2} \epsilon^{i+1,n} + \dfrac{ \sigma_e^{i+1,n} }{2} \dfrac{ 4+\alpha_1 }{6} c_{v,\refe} T^{i+1,n} + \left( \dfrac{ 2-\sigma_j^{i+1,n} }{4} - \dfrac{ 2-\sigma_e^{i+1,n} }{2} \dfrac{ 2-\beta_1 }{12} \right) \delta_t q_c^{i+1,n} . 
    \end{split}
\end{equation}
\end{subequations}

Without loss of generality, the solid-fluid interface is assumed to be located within the lattice cell $i_0$ when the time $n$ is in the $\delta \text{-neighborhood}$ of $n_0$, denoted by $U (n_0, \delta) = \big\{ n \,|\, n_0 - \delta <n< n_0 + \delta \big\}$. Here, the width of $U(n_0, \delta)$ is set to be relatively large to ensure that the interface lattice cell remains $i_0$ in the following discussion. As illustrated by Fig.\ \ref{fig.16}, $i \leq i_0 - 1$ and $i \geq i_0 + 1$ are the solid and fluid lattice cells, respectively. First, the solid lattice cell $i \leq i_0 -1$ is considered, where $f_s = 1$, $T = T_0$, and $\epsilon = \epsilon_s$. In addition, for the solid lattice cell, the relaxation matrix is denoted by $\bfR_s$, and the source term $q_c^{} = -c_v \bfu \cdot \nabla T + (c_{v,s} - c_{v,f}) T \partial f_s \big/ \partial t$ vanishes because there are $\nabla T = \mathbf{0}$ and $\partial f_s \big/ \partial t = 0$. With these conditions, and for $i \leq i_0 - 1$, Eq.\ (\ref{eq.g024}) can be reformulated as 
\begin{equation} \label{eq.g024.s.t1}
    g_{0+2+4}^{i,n+1} - \left( \epsilon_s - \dfrac{4+\alpha_1}{6} c_{v,\refe} T_0 \right) = \left( 1 - \sigma_{e,s} \right) \left[ g_{0+2+4}^{i,n} - \left( \epsilon_s - \dfrac{4+\alpha_1}{6} c_{v,\refe} T_0 \right) \right].
\end{equation}
Recursively substituting Eq.\ (\ref{eq.g024.s.t1}) into itself by varying the time $n$, and considering $|1-\sigma_e| < 1$, Eq.\ (\ref{eq.g024.s.t1}) immediately leads to 
\begin{equation} \label{eq.g024.s}
    g_{0+2+4}^{i,n} = \epsilon_s - \dfrac{4+\alpha_1}{6} c_{v,\refe} T_0 , \qquad \forall\; i \leq i_0 - 1 .
\end{equation}
Therefore, the summation $g_{0+2+4}^{}$ remains constant for the solid lattice cell. For $i \leq i_0$, Eq.\ (\ref{eq.g158}) can be reformulated as
\begin{equation} \label{eq.g158.s.t1}
    g_{1+5+8}^{i,n+1} - \dfrac{ 4+\alpha_1 }{12} c_{v,\refe} T_0 = \left( 1 - \sigma_{j,s} \right) \left( g_{1+5+8}^{i-1,n} - \dfrac{ 4+\alpha_1 }{12} c_{v,\refe} T_0 \right).
\end{equation} 
Recursively substituting Eq.\ (\ref{eq.g158.s.t1}) into itself by varying the time $n$, and considering $|1-\sigma_j| < 1$, Eq.\ (\ref{eq.g158.s.t1}) immediately leads to
\begin{equation} \label{eq.g158.s}
    g_{1+5+8}^{i,n} = \dfrac{ 4+\alpha_1 }{12} c_{v,\refe} T_0 , \qquad \forall\; i \leq i_0 . 
\end{equation}
Therefore, the summation $g_{1+5+8}^{}$ remains constant for the solid and interface lattice cells. To ensure the analytical state of the solid (i.e., $\epsilon = \epsilon_s$ and thus $T = T_0$ for $i \leq i_0 - 1$), the summation $g_{3+6+7}^{}$ is expected to be $\tfrac{4+\alpha_1}{12} c_{v,\refe} T_0$ for the solid lattice cell given Eqs.\ (\ref{eq.g024.s}) and (\ref{eq.g158.s}). For $i \leq i_0 - 2$, Eq.\ (\ref{eq.g367}) can be reformulated as
\begin{equation} \label{eq.g367.s.t1}
    g_{3+6+7}^{i,n+1} - \dfrac{4+\alpha_1}{12} c_{v,\refe} T_0 = \left( 1 - \sigma_{j,s} \right) \left( g_{3+6+ 7}^{i+1,n} - \dfrac{4 + \alpha_1}{12} c_{v,\refe} T_0 \right),
\end{equation}
which suggests that the quantity $g_{3+6+7}^{} - \tfrac{4+\alpha_1}{12} c_{v,\refe} T_0$, if it is nonzero, will damply transfer into the solid starting from the rightmost solid lattice cell $i_0 - 1$. Therefore, the rightmost solid lattice cell $i_0 - 1$, closest to the solid-fluid interface, is the critical factor in ensuring the analytical state of the solid. For $i = i_0 -1$, the summation $g_{3+6+7}^{}$ can be calculated from Eq.\ (\ref{eq.g367}) as follows: 
\begin{equation}
    g_{3+6+7}^{i_0 -1,n+1} = \dfrac{4+\alpha_1}{12} c_{v,\refe} T_0 + \dfrac{2-\sigma_e^{i_0, n} }{4} \dfrac{4+\beta_1}{6} \delta_t q_c^{i_0, n} + \dfrac{ \sigma_j^{i_0,n} + \sigma_e^{i_0,n} - 2 }{2} \left( g_{0+2+4}^{i_0,n} - \epsilon^{i_0,n} + \dfrac{4 + \alpha_1}{6} c_{v,\refe} T_0 + \dfrac{1}{2} \delta_t q_c^{i_0,n} \right),
\end{equation}
where the last two terms are expected to be zero. For this purpose, we can set
\begin{subequations} \label{eq.con.s}
    \begin{gather}
        \beta_1 = -4 ,
        \\
        \label{eq.con.s.2}
        \sigma_j^{i_0,n} + \sigma_e^{i_0,n} = 2  \quad \mathrm{or} \quad g_{0+2+4}^{i_0,n} = \epsilon^{i_0,n} - \dfrac{4 + \alpha_1}{6} c_{v,\refe} T_0 - \dfrac{1}{2} \delta_t q_c^{i_0,n} .
    \end{gather}
\end{subequations}

Then, the fluid lattice cell $i \geq i_0 +1$ is considered, where $f_s = 0$, $T = T_0$, and $\epsilon = \epsilon_f$. In addition, the relaxation matrix is denoted by $\bfR_f$, and similar to the solid lattice cell, the source term $q_c^{} = -c_v \bfu \cdot \nabla T + (c_{v,s} - c_{v,f}) T \partial f_s \big/ \partial t$ also vanishes here. With these conditions, and for $i \geq i_0 +1$, Eq.\ (\ref{eq.g024}) can be reformulated as 
\begin{equation} \label{eq.g024.f.t1}
    g_{0+2+4}^{i,n+1} - \left( \epsilon_f - \dfrac{ 4 + \alpha_1 }{6} c_{v,\refe} T_0 \right) = \left( 1 - \sigma_{e,f} \right) \left[ g_{0+2+4}^{i,n} - \left( \epsilon_f - \dfrac{ 4 + \alpha_1 }{6} c_{v,\refe} T_0 \right) \right] .
\end{equation}
Recursively substituting Eq.\ (\ref{eq.g024.f.t1}) into itself by varying the time $n$, and considering $| 1-\sigma_e | < 1$, Eq.\ (\ref{eq.g024.f.t1}) immediately leads to 
\begin{equation} \label{eq.g024.f}
    g_{0+2+4}^{i,n} = \epsilon_f - \dfrac{ 4 + \alpha_1 }{6} c_{v,\refe} T_0 , \qquad \forall\; i \geq i_0 + 1. 
\end{equation}
Therefore, the summation $g_{0+2+4}^{}$ remains constant for the fluid lattice cell. For $i \geq i_0$, Eq.\ (\ref{eq.g367}) can be reformulated as 
\begin{equation} \label{eq.g367.f.t1}
    g_{3+6+7}^{i,n+1} - \dfrac{4 + \alpha_1}{12} c_{v,\refe} T_0  =  \left( 1 - \sigma_{j,f} \right) \left( g_{3+6+7}^{i+1,n} - \dfrac{4 + \alpha_1}{12} c_{v,\refe} T_0 \right).
\end{equation}
Recursively substituting Eq.\ (\ref{eq.g367.f.t1}) into itself by varying the time $n$, and considering $| 1-\sigma_j | <1$, Eq.\ (\ref{eq.g367.f.t1}) immediately leads to 
\begin{equation} \label{eq.g367.f}
    g_{3+6+7}^{i,n} = \dfrac{4 + \alpha_1}{12} c_{v,\refe} T_0 , \qquad \forall\; i \geq i_0 .
\end{equation}
Therefore, the summation $g_{3+6+7}^{}$ remains constant for the fluid and interface lattice cells. To ensure the analytical state of the fluid (i.e., $\epsilon = \epsilon_f$ and thus $T = T_0$ for $i \geq i_0 +1$), the summation $g_{1+5+8}^{}$ is expected to be $\tfrac{4 + \alpha_1}{12} c_{v,\refe} T_0$ for the fluid lattice cell given Eqs.\ (\ref{eq.g024.f}) and (\ref{eq.g367.f}). For $i \geq i_0 +2$, Eq.\ (\ref{eq.g158}) can be reformulated as
\begin{equation} \label{eq.g158.f.t1}
    g_{1+5+8}^{i,n+1} - \dfrac{4 + \alpha_1}{12} c_{v,\refe} T_0 = \left( 1 - \sigma_{j,f} \right) \left( g_{1+5+8}^{i-1,n} - \dfrac{ 4 + \alpha_1 }{12} c_{v,\refe} T_0 \right) ,
\end{equation}
which suggests that the quantity $g_{1+5+8}^{} - \tfrac{4 + \alpha_1}{12} c_{v,\refe} T_0$, if it is nonzero, will damply transfer into the fluid starting from the leftmost fluid lattice cell $i_0 +1$. Therefore, the leftmost fluid lattice cell $i_0 +1$, closest to the solid-fluid interface, is the critical factor in ensuring the analytical state of the fluid. For $i = i_0+1$, the summation $g_{1+5+8}^{}$ can be calculated from Eq.\ (\ref{eq.g158}) as follows: 
\begin{equation}
    g_{1+5+8}^{i_0 + 1,n + 1} = \dfrac{ 4 + \alpha_1 }{12} c_{v,\refe} T_0 + \dfrac{ 2 - \sigma_e^{i_0,n} }{4} \dfrac{ 4 + \beta_1 }{6} \delta_t q_c^{i_0,n} + \dfrac{ \sigma_j^{i_0,n} + \sigma_e^{i_0,n} - 2 }{2} \left( g_{0+2+4}^{i_0,n} - \epsilon^{i_0,n} + \dfrac{4 + \alpha_1}{6} c_{v,\refe} T_0 + \dfrac{1}{2} \delta_t q_c^{i_0,n} \right) ,
\end{equation}
where the last two terms are expected to be zero. For this purpose, we can set 
\begin{subequations} \label{eq.con.f}
    \begin{gather}
        \beta_1 = -4 ,
        \\
        \label{eq.con.f.2}
        \sigma_j^{i_0,n} + \sigma_e^{i_0,n} = 2  \quad \mathrm{or} \quad g_{0+2+4}^{i_0,n} = \epsilon^{i_0,n} - \dfrac{4 + \alpha_1}{6} c_{v,\refe} T_0 - \dfrac{1}{2} \delta_t q_c^{i_0,n} .
    \end{gather}
\end{subequations}
Here, it is worth pointing out that Eq.\ (\ref{eq.con.f}) is identical to Eq.\ (\ref{eq.con.s}), which means that the analytical states of the solid and fluid are achieved, or not, at the same time.

At last, the interface lattice cell $i = i_0$ is considered, where $T = T_0$ but $f_s$ and $\epsilon$ vary with time. In addition, the relaxation matrix also varies with time and is denoted by $\bfR^{i_0, n}$, and the source term $q_c^{} = -c_v \bfu \cdot \nabla T + (c_{v,s} - c_{v,f}) T \partial f_s \big/ \partial t$ is nonzero and can be simplified as $q_c^{i_0,n} = (c_{v,s} - c_{v,f}) T_0 \big( \partial f_s \big/ \partial t \big) ^{i_0, n}$ because $\nabla T = \mathbf{0}$, $c_{v,s} \neq c_{v,f}$, and $\partial f_s \big/ \partial t \neq 0$. It can be known from Eqs.\ (\ref{eq.g158.s}) and (\ref{eq.g367.f}) that, for the interface lattice cell, there are 
\begin{equation}
    g_{1+5+8}^{i_0,n} = g_{3+6+7}^{i_0,n} = \dfrac{4 + \alpha_1}{12} c_{v,\refe} T_0 .
\end{equation}
Considering the definition $\epsilon = \sum\nolimits_{i=0}^8 g_i^{} + \tfrac{\delta_t}{2} q_c^{}$ and to ensure $T^{i_0,n} = T_0$, the summation $g_{0+2+4}^{i_0,n}$ is expected to be $\epsilon^{i_0,n} - \tfrac{4 + \alpha_1}{6} c_{v,\refe} T_0 - \tfrac{1}{2} \delta_t q_c^{i_0,n}$ with $\epsilon^{i_0,n} = c_v^{i_0,n} T_0 = [c_{v,f} + (c_{v,s} - c_{v,f}) f_s^{i_0,n}] T_0$. For $i = i_0$, Eq.\ (\ref{eq.g024}) can be reformulated as
\begin{equation}
    \begin{split}
        g_{0+2+4}^{i_0, n+1} - \epsilon^{i_0,n+1} + \dfrac{4 + \alpha_1}{6} c_{v,\refe} T_0 + \dfrac{1}{2} \delta_t q_c^{i_0,n+1} =  {} & \left( 1 - \sigma_e^{i_0,n} \right) \left( g_{0+2+4}^{i_0,n} - \epsilon^{i_0,n} + \dfrac{4 + \alpha_1}{6} c_{v,\refe} T_0 + \dfrac{1}{2} \delta_t q_c^{i_0,n} \right) \\ 
        {} & - \dfrac{2 - \sigma_e^{i_0,n} }{2} \dfrac{4 + \beta_1}{6} \delta_t q_c^{i_0,n} - \left( \epsilon^{i_0,n + 1} - \epsilon^{i_0,n} - \delta_t \dfrac{ q_c^{i_0,n + 1} + q_c^{i_0,n} }{2} \right) , 
    \end{split}
\end{equation}
where the last two terms on the right-hand side are expected to be zero. For this purpose, we can set 
\begin{subequations} \label{eq.con.i}
    \begin{gather}
        \beta_1 = -4 ,
        \\
        \label{eq.con.i.2}
        \epsilon^{i_0,n + 1} - \epsilon^{i_0,n} = \delta_t \dfrac{ q_c^{i_0,n + 1} + q_c^{i_0,n} }{2} .
    \end{gather}
\end{subequations}
Substituting $\epsilon^{i_0,n} = [c_{v,f} + (c_{v,s} - c_{v,f}) f_s^{i_0,n}] T_0$ and $q_c^{i_0,n} = (c_{v,s} - c_{v,f}) T_0 \big( \partial f_s \big/ \partial t \big) ^{i_0, n}$ into Eq.\ (\ref{eq.con.i.2}), we finally have 
\begin{equation} \label{eq.con.dfs}
    \left( \dfrac{ \partial f_s }{ \partial t } \right)^{i_0, n + 1} + \left( \dfrac{ \partial f_s }{ \partial t } \right)^{i_0,n} = 2 \dfrac{ f_s^{i_0, n + 1} - f_s^{i_0,n} }{\delta_t} ,
\end{equation}
which acts as a constraint in computing the time derivative $\partial f_s \big/ \partial t$ in real simulations. Note that the expected $g_{0+2+4}^{i_0,n} = \epsilon^{i_0,n} - \tfrac{4 + \alpha_1}{6} c_{v,\refe} T_0 - \tfrac{1}{2} \delta_t q_c^{i_0,n}$ for the interface lattice cell is precisely the second relation in Eqs.\ (\ref{eq.con.s.2}) and (\ref{eq.con.f.2}), implying that the first relation $\sigma_j^{i_0,n} + \sigma_e^{i_0,n} = 2$ is not required. Based on the above analysis, it can be concluded that only $\beta_1 = -4$ and Eq.\ (\ref{eq.con.dfs}) are necessary to ensure the analytical state of the solid-fluid system. Otherwise, temperature deviation will be induced at the interface lattice cell, which damply transfers into the solid and fluid as described by Eqs.\ (\ref{eq.g367.s.t1}) and (\ref{eq.g158.f.t1}), respectively.

%

\end{document}